

\input harvmac

\def\sm{$S$-matrix}\def\sms{$S$-matrices}

\def\scamps{scattering amplitudes}
\def\smt{$S$-matrix theory}
\def\sme{$S$-matrix element}\def\smes{$S$-matrix elements}
\def\pf{partition function}
\def\pfs{partition functions}
\def\lag{lagrangian}
\def\ZrmSG{$Z_{{\rm SG}}$}
\def\SGf{SG$(\sqrt{4\pi})$}
\def\s4p{\sqrt{4\pi}}

\def\ie{{\it i.e.}}
\def\eg{{\it e.g.}}
\def\no{\noindent}
\def\o{\over}
\def\ra{\rangle}\def\la{\langle}

\def\scsc{\scriptscriptstyle}
\def\sN{{\scriptscriptstyle N}}

\def\nl{\hfill\break}
\def\alp{\alpha}\def\gam{\gamma}\def\del{\delta}\def\lam{\lambda}
\def\eps{\epsilon}\def\veps{\varepsilon}
\def\Del{\Delta}
\def\sig{\sigma}\def\th{\theta}

\def\r{\rho}
\def\rb{{\rm b}}
\def\rf{{\rm f}}
\def\ah{\hat{a}}
\def\zb{\bar{z}}
\def\wb{\bar{w}}

\def\qb{\bar{q}}
\def\pb{\bar{\partial}}
\def\sb{\bar{s}}
\def\phib{\bar{\phi}}
\def\Phit{\tilde{\Phi}}
\def\Psib{\bar{\Psi}}

\def\Ot{\tilde{O}}\def\Ut{\tilde{U}}
\def\Delb{\bar{\Delta}}

\def\ZZ{{\bf Z}}
\def\NN{{\bf N}}
\def\RR{{\bf R}}

\def\ontopss#1#2#3#4{\raise#4ex \hbox{#1}\mkern-#3mu {#2}}

\nref\rFroSG{J.~Fr\"ohlich, in: {\it Renormalization Theory} (Erice 1975),
  ed. G.~Velo and A.S.~Wightman (Reidel, Dordrecht, 1976); ~J.~Fr\"ohlich
 and E.~Seiler, Helv.~Phys.~Acta~49 (1976) 889}
\nref\rThir{W.~Thirring, Ann.~Phys.~(NY) 3 (1958) 91}
\nref\rGlas{V.~Glaser, Nuovo Cimento~9 (1958) 990}
\nref\rJohn{K.~Johnson, Nuovo Cimento~20 (1961) 773}
\nref\rWigh{A.S.~Wightman, in: {\it Carg\`ese Lectures in Theoretical Physics},
 1964, ed. M. Levy (Gordon and Breach, New York, 1966)}
\nref\rKlai{B.~Klaiber, in: {\it Lectures in Theoretical Physics} (Boulder
 1967), ed.~A.~Barut and W.~Brittin (Gordon and Breach, New York, 1968),
 Vol.~X, part A}
\nref\rDA{G.F.~Dell'Antonio, Acta Phys.~Austriaca 43 (1975) 43}
\nref\rCole{S.~Coleman, Phys.~Rev.~D11 (1975) 2088}
\nref\rMand{S.~Mandelstam, Phys.~Rev.~D11 (1975) 3026}
\nref\rBPZ{A.A.~Belavin, A.M.~Polyakov and A.B.~Zamolodchikov,
 Nucl.~Phys.~B241 (1984) 333}
\nref\rCarLH{J.L.~Cardy, in: {\it Fields, Strings, and Critical Phenomena},
 Les Houches 1988,
 ed. E.~Br\'ezin and J.~Zinn-Justin, (North Holland, Amsterdam, 1989)}
\nref\rGinspLH{P.~Ginsparg, in: {\it Fields, Strings, and Critical Phenomena},
 Les Houches 1988,
 ed. E.~Br\'ezin and J.~Zinn-Justin, (North Holland, Amsterdam, 1989)}
\nref\rMICar{J.L.~Cardy, Nucl.~Phys.~B270 (1986) 186;
 {\it ibid.}~B275 (1986) 200}
\nref\rCIZ{A.~Cappelli, C.~Itzykson and J.-B.~Zuber, Nucl.~Phys.~B280
 (1987) 445}
\nref\rGepn{D.~Gepner, Nucl.~Phys.~B287 (1987) 111}
\nref\rFenFSM{P.~Fendley, Phys.~Lett.~250B (1990) 96}
\nref\rFenInt{P.~Fendley and K.~Intriligator,~Nucl.~Phys.~B372~(1992)~533,
 ~``Scattering and Thermodynamics in
 Integrable $N=2$ Theories'', BUHEP-92-5/HUTP-91-A067 (1992)}
\nref\rFS{D.~Friedan and S.~Shenker, in: {\it Conformal Invariance and
 Applications to Statistical Mechanics}, ed.~C.~Itzykson {\it et al}
 (World Scientific, Singapore, 1988), p.~578}
\nref\rDFMS{L.~Dixon, D.~Friedan, E.~Martinec and S.H.~Shenker,
 Nucl.~Phys.~B282 (1987) 13}
\nref\rGSO{F.~Gliozzi, J.~Scherk and D.~Olive, Nucl.~Phys.~B122 (1977) 253}
\nref\rZams{A.B.~Zamolodchikov and Al.B.~Zamolodchikov, Ann.~Phys.~(NY)~120
 (1980) 253}
\nref\rKarThu{M.~Karowski and H.J.~Thun, Nucl.~Phys.~B130 (1977) 295}
\nref\rKore{V.E.~Korepin, Teor.~Mat.~Fiz.~41 (1979) 169}
\nref\rBeLe{D.~Bernard and A.~LeClair, Commun.~Math.~Phys.~142 (1991) 99}
\nref\rKar{M.~Karowski, Nucl.~Phys.~B153 (1979) 244}
\nref\rZamtba{Al.B.~Zamolodchikov, Nucl.~Phys.~B342 (1990) 695}
\nref\rouriv{T.R.~Klassen and E.~Melzer, Nucl.~Phys.~B362 (1991) 329}
\nref\rourviii{T.R.~Klassen and E.~Melzer, ``Kinks in Finite Volume'',
 Cornell/Stony Brook preprint CLNS-92-1130/ITP-SB-92-01, to appear in
 Nucl.~Phys.~B}
\nref\rKir{E.B.~Kiritsis, Phys.~Lett.~217B (1989) 427}
\nref\rFP{D.Z.~Freedman and K.~Pilch, Phys.~Lett.~213B (1988) 331,
 and Ann.~Phys. (NY) 192 (1989) 331}
\nref\rDAFZ{G.F.~Dell'Antonio, Y.~Frishman and D.~Zwanziger, Phys.~Rev.~D6
 (1972) 988}
\nref\rLuth{A.~Luther, Phys.~Rev.~B14 (1976) 2153}
\nref\rThST{H.~Neuberger, A.J.~Niemi and G.W.~Semenoff, Phys.~Lett.~181B
 (1986) 244; H.~Kawai, D.C.~Lewellen and S.-H.~Tye, Nucl.~Phys.~B288 (1987) 1;
 J.~Bagger, D.~Nemeschansky, N.~Seiberg and S.~Yankielowicz, Nucl.~Phys.~B289
 (1987) 53}
\nref\rsFQS{D.~Friedan, Z.~Qiu and S.H.~Shenker, Phys.~Lett.~151B (1985) 37}
\nref\rZamflow{A.B.~Zamolodchikov, Sov.~J.~Nucl.~Phys.~46 (1987) 1090}
\nref\rLudCar{A.W.W.~Ludwig and J.L.~Cardy, Nucl.~Phys.~B285 (1987) 687}
\nref\rTanig{A.B.~Zamolodchikov, Adv.~Stud.~Pure Math.~19 (1989) 1}
\nref\rHenSal{M.~Henkel and H.~Saleur, J.~Phys.~A23 (1990) 791}
\nref\rournext{T.R.~Klassen and E.~Melzer, in preparation}
\nref\rSmirBook{F.A.~Smirnov, {\it Form Factors in Completely Integrable
Models of Quantum Field Theory} (World Scientific, Singapore, 1992)}
\nref\rKTTW{M.~Karowski, H.J.~Thun, T.T.~Truong, P.~Weisz, Phys.~Lett.~67B
(1977) 321}
\nref\rWeisz{P.~Weisz, Nucl.~Phys.~B122 (1977) 1}
\nref\rBjDr{J.~Bjorken and S.~Drell, {\it Relativistic Quantum Fields}
           (McGraw-Hill, New York, 1965)}
\nref\rourii{T.R.~Klassen and E.~Melzer, Nucl.~Phys.~B338 (1990) 485}
\nref\rLMC{M.~L\"assig, G.~Mussardo and J.L.~Cardy, Nucl.~Phys.~B348 (1991)
 591}
\nref\rouriii{T.R.~Klassen and E.~Melzer, Nucl.~Phys.~B350 (1991) 635}
\nref\rFenExc{P.~Fendley, Nucl. Phys. B374 (1992) 667}
\nref\rKMS{D.A.~Kastor, E.J.~Martinec and S.H.~Shenker,
 Nucl.~Phys.~B316 (1989) 590}
\nref\rFSSC{J.L.~Cardy, J.~Phys.~A17 (1984) L385;
 H.W.J.Bl\"ote, J.L.~Cardy and M.P.~Nightingale, Phys.~Rev.~Lett.
 56 (1986) 742;
 I.~Affleck, Phys.~Rev.~Lett. 56 (1986) 746}
\nref\rYZ{V.P.~Yurov and Al.B.~Zamolodchikov, Int.~J.~Mod.~Phys.~A5 (1990)
 3221, and Paris preprint ENS-LPS-273 (1990)}
\nref\rLasMus{M.~L\"assig and G.~Mussardo, Computer Phys.~Comm.~66 (1991)
 71}
\nref\rourv{T.R.~Klassen and E.~Melzer, Nucl.~Phys.~B370 (1992) 511}
\nref\rIFT{T.T.~Wu, B.M.~McCoy, C.A.~Tracy and E.~Baruch, Phys.~Rev.~B13
 (1976) 316;\nl
 B.~Schroer and T.T.~Truong, Nucl.~Phys.~B144 (1978) 80}
\nref\rSMJ{M.~Sato, T.~Miwa and M.~Jimbo, Proc.~Japan Acad.~53A (1977)
 6, 147, 153, 183, 219}
\nref\rFerFi{A.E.~Ferdinand and M.E.~Fisher, Phys.~Rev.~185 (1969) 832}
\nref\rSalItz{H.~Saleur and C.~Itzykson, J.~Stat.~Phys.~48 (1987) 449}
\nref\rLui{M.~L\"uscher, in: {\it Progress in Gauge Field Theory}
 (Carg\`ese 1983), ed. G.~'t~Hooft {\it et al} (Plenum, New York, 1984), and
 Commun.~Math.~Phys.~104 (1986) 177}
\nref\rLuii{M.~L\"uscher, Commun.~Math.~Phys.~105 (1986) 153, and
  Nucl.~Phys.~B354 (1991) 531;
 M.~L\"uscher and U.~Wolff, Nucl.~Phys.~B339 (1990) 222}
\nref\rLasMar{M.~L\"assig and M.J.~Martins, Nucl.~Phys.~B354 (1991)
 666;\nl M.J.~Martins, Phys.~Lett.~262B (1991) 39}
\nref\rbeta{ I.M.~Gel'fand, M.I.~Graev and I.I.~Pyatetskii-Shapiro,
 {\it Representation The\-ory and Auto\-mo\-rphic Func\-tions}
  (Saunders, Philadelphia,
 1966);  ~E.~Melzer, Int.~J.~Mod.~Phys. A4 (1989) 4877}
\nref\rSchur{M.~Wakimoto and H.Yamada, Lett.~Math.~Phys.~7 (1983) 513}
\nref\rWZWFZ{E.~Witten, Commun.~Math.~Phys.~92 (1984) 455; \nl
 A.B.~Zamolodchikov and V.A.~Fateev, Sov.~J.~Nucl.~Phys.~43 (1986) 657}
\nref\rColeSchw{S.~Coleman, Ann.~Phys.~(NY)~101~(1976)~239}
\nref\rMWsusy{S.~Meyer and P.~Weisz, Phys.~Lett.~68B~(1977)~471}
\nref\rAraKor{I.Ya.~Aref'eva and V.E.~Korepin, JETP Lett.~20 (1974) 312}
\nref\rDaHaNe{R.F.~Dashen, B.~Hasslacher and A.~Neveu, Phys.~Rev.~D10 (1974)
   4114, 4130, ~D11 (1975) 3424, 2443}
\nref\rColeBook{S.~Coleman, {\it Aspects of Symmetry} (Cambridge University
  Press, Cambridge, 1985)}
\nref\rZamDiv{A.B.~Zamolodchikov, Commun.~Math.~Phys.~69 (1979) 155}
\nref\rFenGin{P.~Fendley and P.~Ginsparg, Nucl.~Phys.~B324~(1989)~549}
\nref\rZamkink{A.B.~Zamolodchikov, Landau Institute preprint (1989)}
\nref\rHenLud{M.~Henkel and A.W.W.~Ludwig, {\it Mass Spectrum of the 2D
  Ashkin-Teller Model in an External Magnetic Field}, Geneva preprint
  UGVA/DPT 1990/05-672}
\nref\rDVV{R.~Dijkgraaf, E.~Verlinde and H.~Verlinde, in: {\it Perspectives
 in String Theory}, ed.~P.~Di Vecchia and J.L.~Petersen (World Scientific,
 Singapore, 1988)}
\nref\rTB{T.~Banks, in: {\it The Santa Fe TASI-87}, ed.~R.~Slansky and G.~West
   (World Scientific, Singapore, 1988)}
\nref\rNahm{W.~Eholzer, M.~Flohr, A.~Honecker, R.~H\"ubel, W.~Nahm,
 and R.~Varnhagen, Bonn preprint BONN-HE-91-22, and references therein}
\nref\rourvii{T.R.~Klassen and E.~Melzer, ``RG Flows in the $D$-Series of
 Minimal CFTs'', Cornell/Stony Brook preprint CLNS-91-1111/ITP-SB-91-57
 (1991)}
\nref\rKiZh{Q.~Ho-Kim and H.B.~Zheng, Phys.~Lett.~212B (1988) 71}
\nref\rRSG{A.~LeClair, Phys.~Lett.~230B (1989) 103;\nl
  D.~Bernard and A.~LeClair, Nucl.~Phys.~B340 (1990) 721;\nl
  N.Yu.~Reshetikhin and F.A.~Smirnov, Commun.~Math.~Phys.~131 (1990) 157}
\nref\rXXZi{S.V.~Pokrovski and A.M.~Tsvelik, Sov.~Phys.~JETP 66 (1987) 1275;
 \nl
 F.C.~Alcaraz, M.~Baake, U.~Grimm and V.~Rittenberg, J.~Phys.~A21 (1988) L117}
\nref\rXXZii{A.~Luther and I.~Peschel, Phys.~Rev.~B12 (1975) 3908; \nl
 V.E.~Korepin and A.G.~Izergin, JETP Lett.~42 (1985) 512}
\nref\rIsnum{M.~Henkel and H.~Saleur, J.~Phys.~A22 (1989) L513;
 I.R.~Sagdeev and A.B.~Zamolodchikov, Mod.~Phys.~Lett.~B3 (1989) 1375;
 P.G.~Lauwers and V.~Rittenberg, Phys.~Lett. 233B (1989) 197;
 M.~Henkel, J.~Phys.~A24 (1991) L133}
\nref\rBaxt{R.J.~Baxter, {\it Exactly Solved Models in Statistical mechanics}
 (Academic Press, London, 1982)}
\nref\rLuMTM{M.~L\"uscher, Nucl.~Phys.~B117 (1976) 475}
\nref\rKauf{B.~Kaufman, Phys.~Rev.~76 (1949) 1232}

\Title{\vbox{\baselineskip12pt\hbox{CLNS-92/1149~~ITP-SB-92-36}
\hbox{hep-th@xxx/9206114} }}
{\vbox{\centerline{Sine-Gordon $\neq$ Massive Thirring,}
\vskip6pt\centerline{and Related Heresies}}}

\centerline{Timothy R.~Klassen\foot{{\it Newman Laboratory,
  Cornell University,  Ithaca, NY 14853} ~ klassen@strange.tn.cornell.edu}
{}~ and ~ Ezer Melzer\foot{{\it Inst.~for~Theor.~Physics,
      SUNY, Stony Brook,  NY~11794} ~ melzer@max.physics.sunysb.edu}}

\vskip 5mm
By viewing the Sine-Gordon and massive Thirring models as perturbed
conformal field theories one sees that they are different (the difference
being observable, for instance, in finite-volume energy levels).
The UV limit of the former (SGM) is a gaussian model,
that of the latter (MTM)  a so-called {\it fermionic} gaussian model,
the compactification radius of the boson underlying both theories
depending on     the SG/MT coupling.
(These two families of conformal field theories are related by
a ``twist''.)
Corresponding SG and MT models contain a subset of fields with identical
correlation functions, but  each model also has fields the other one does not,
{\it e.g.}~the fermion fields of MTM are not contained in SGM, and the
{\it bosonic} soliton fields of SGM are not in MTM.
Our results imply, in particular, that the SGM at the so-called
``free-Dirac point'' $\beta^2\!=\!4\pi$ is actually a theory of two
interacting bosons with diagonal $S$-matrix $S=-{\sl 1}$, and that for
arbitrary couplings the overall sign of the accepted SG $S$-matrix in the
soliton sector should be reversed.
{}~More generally, we draw attention to the existence of  new
classes of quantum field theories, analogs of the (perturbed) fermionic
gaussian models, whose partition functions are invariant only under a
subgroup of the modular group. One such class comprises
``fermionic versions'' of the Virasoro minimal models.

\Date{\hfill 6/92}

\vfill\eject

\newsec{Introduction}
\ftno=0

The sine-Gordon model (SGM) is a (1+1)-dimensional field theory of a
pseudo-scalar field $\varphi$, defined classically by the lagrangian
\eqn\lagSG{ {\cal L}_{{\rm SG}} ~=~
    {1\o 2}\partial_\mu \varphi \partial^\mu \varphi
      ~-~ {\alp_0\o \beta^2}~(1- \cos\beta\varphi) ~.}
Here $\alp_0$ is a mass scale, $\beta$ a dimensionless coupling, and
one identifies field configurations that differ by a
period ${2\pi\o \beta}$
 of the potential. It has been shown rigorously~\rFroSG\ that one can make
sense out of this theory also on
the quantum level
(at least for a certain range of $\beta$),
the well-known classical
(multi-)soliton solutions of~\lagSG\
corresponding
to nontrivial super-selection sectors in the quantum theory.

The massive Thirring model (MTM) is formally defined by the \lag\
\eqn\lagMTM{ {\cal L}_{{\rm MTM}} ~=~i\bar{\Psi} \rlap/\partial \Psi
      ~-~ m_0 \bar{\Psi} \Psi ~-~ {g \o 2} J_\mu J^\mu ~~~, }
where $J^\mu =  \bar{\Psi} \gam^\mu \Psi$, in terms of a Dirac field $\Psi$.
The quantum theory for the massless case $m_0=0$, the Thirring model, was
proposed in~\rThir, and discussed with increasing sophistication
in~\rGlas\rJohn\rWigh, arbitrary Green's functions of $\Psi$ finally being
written down in~\rKlai.
The {\it a priori} ill-defined product of operators appearing
in the current $J^\mu$ can be defined
by requiring $J^\mu$ to obey
appropriate Ward identities. There is
(at least)
a 1-parameter family of
definitions of $J^\mu$ and one must be careful to specify which is used,
otherwise the dimensionless
coupling $g$ in~\lagMTM\ has no meaning.\foot{Below
and in sect.~2 we will see that the convention-independent way of defining
the coupling in the Thirring model is in terms of the compactification radius
$r$ of a free massless (pseudo-)scalar field.}
It is advantageous to view the
{\it massive} Thirring model
as a perturbation of the massless one~\rDA\rCole\    by the
(suitably regularized) operator $\Psib \Psi$,
           rather than a perturbation
of a free massive           Dirac theory    by $J_\mu J^\mu$,
in the same way as one
can attempt
to define the SGM as a perturbation of its UV limit
$\alp_0=0$
by  $\cos\beta\varphi$~\rCole\rFroSG.
This general idea, defining a (1+1)-dimenional massive quantum field theory
(QFT) as a perturbation of the conformal field theory (CFT) describing
its UV limit, has been very successful in recent years.
This approach is now known as {\it conformal perturbation theory} (CPT);
it will be briefly discussed in sect.~4.2.

\medskip
Since the
work of Coleman~\rCole\ and Mandelstam~\rMand\ it is widely believed
that the SGM and the MTM are equivalent, just being different \lag\
representations of the same underlying QFT, \ie~that there is a 1$-$1 map
between operators in the SGM and the MTM such that corresponding correlation
functions are identical.
This is not what is proved in~\rCole\rMand.
Instead,  Coleman basically
showed that the correlation functions of the perturbing
fields  $\Psib \Psi$ and $\cos\beta\varphi$, respectively, are identical
in the MT and SG models,
provided their couplings are related by
\eqn\Colreln{ 1 ~+~ {g\o \pi} ~=~ {4\pi\o \beta^2} ~,}
in Coleman's conventions. The proof is given to all orders of CPT where it
amounts to showing that all $N$-point functions of the
perturbing fields
are identical in the {\it massless} theories.
Mandelstam showed how to construct a fermion operator satisfying the MTM
equation as a nonlocal functional of a
pseudo-scalar field satisfying the
SG equation. This is done directly in the massive theory.

\medskip
Let us first address Coleman's results.
In the wake of~\rBPZ\ many classes of  CFTs have been understood in great
detail (see~\rCarLH\rGinspLH\ for reviews and references).
There are numerous examples of       distinct CFTs which nevertheless
share a nontrivial subalgebra of operators.
The most familiar examples are Virasoro minimal CFTs~\rBPZ\
of the same central charge $c$ that belong to different series of
modular invariants~\rMICar\rCIZ\rGepn.
Perturbations of two such
CFTs by a suitable common operator will
lead to massive QFTs whose correlation functions are absolutely identical
for certain fields, but different for others. A well-understood  example,
that will prove quite analogous to the case of MTM versus SGM, is that of the
free Majorana fermion versus the Ising field theory, to be discussed in
sect.~4.3.
Other
examples, involving local versus nonlocal realizations of
supersymmetry, were recently discussed in~\rFenFSM\rFenInt.

\medskip
So to investigate whether the SG and MT models
really are   equivalent it is useful
to first study the CFTs describing their UV limits.
Both models have an $O(2)$$\times$$O(2)$
internal symmetry in the massless limit, so their
Hilbert spaces will split into super-selection  sectors corresponding to the
inequivalent representations of the $U(1)$$\times$$U(1)$
current algebra.
These sectors are created by vertex operators, which can be expressed in terms
of the left- and right-moving
parts of a compactified massless scalar field. As we will discuss   in detail
in sect.~2, for a given compactification radius $r$ there
are exactly two maximal, closed sets of mutually local vertex operators. One
corresponds to  the ``standard'' gaussian model,
the other to the ``fermionic gaussian model''~\rFS.
These two families of CFTs have central charge $c$=1 for all $r$,
and are related through
a so-called $\ZZ_2$ twist~\rDFMS\ or GSO projection~\rGSO.
The vertex operators $V_{m,n}$  are labeled by
``electric'' and ``magnetic'' charges, $m,n$, respectively. In the standard,
or {\it bosonic} gaussian model, $m,n\in \ZZ$, whereas in the fermionic one
$m\in 2\ZZ, n\in \ZZ$ or $m\in 2\ZZ +1, n\in \ZZ+{1\o 2}$.
With  our conventions the Lorentz spin of $V_{m,n}$ is $m n$.
All of this
is the content of sect.~2.

The UV limit of the SGM is
the bosonic gaussian model.  The MTM, on
the other hand, contains fermionic fields, \ie~fields of half-odd-integer
Lorentz spin which anti-commute at spacelike separations.
Therefore
it can only
correspond to the fermionic gaussian model      in the UV limit.

We see that the bosonic and fermionic gaussian models have a common closed
subset of sectors, labeled by $m\in 2\ZZ, n\in \ZZ$. In particular, both
CFTs contain the field $(V_{0,1}+V_{0,-1}) \propto \cos\beta\varphi$.
Perturbing by this field gives the SGM and MTM, respectively,
as will be discussed in sect.~3.
The fields corresponding to $m\in 2\ZZ, n\in \ZZ$
have identical correlation functions also in the massive theory
(to all orders of CPT).
The result of~\rCole\ is a special case of this;
 in our notation, Coleman showed that correlators
 involving the same number of  massive analogs of
 $V_{0,1}$ and $V_{0,-1}$ are identical in SGM and MTM with
 the same compactification radius of the UV boson.

But as we will see  in sect.~3,
the solitons in SGM are created
by massive analogs of $V_{\pm 1,0}$, whereas the bosonized
components of the Dirac
field $\Psi$ ($\Psi^\dagger$) creating the fermions of MTM are the
massive versions of $V_{1,\pm {1\o 2}}$ ~($V_{-1,\mp {1\o 2}}$).
All these operators carry $\pm 1$ unit of
   the
$U(1)$ charge
   that remains conserved in the massive theories,
but have
Lorentz spins $0$ and $\pm {1\o 2}$ in the respective models,
showing
that the SG solitons are {\it bosons} which can not
be identified with the MT fermions!

Since in both the bosonic and fermionic gaussian models all operators can be
expressed in terms of ``the same''
(pseodo-)scalar
field $\varphi$, and its massive analog obeys the SG equation
in both perturbed theories, it is not
implausible that also the operators in the massive theories can be expressed in
terms of the massive $\varphi$.\foot{Note
  that $\varphi$ itself is not an operator in SGM or MTM, as we will see,
but serves as
``building block'' of the theories, and under closer inspection
will actually be seen to have different periodicity properties in SGM and MTM.}
Indeed,
Mandelstam showed~\rMand\ that the Dirac field $\Psi$ of the
MTM can be written (nonlocally) in terms of $\varphi$.
However, this has no bearing on the question whether the SG and MT
models are equivalent.
Mandelstam's work shows just as well
(naively it is actually more obvious)
that the bosonic fields which we identify as creating the solitons of the SGM
can be written in terms of $\varphi$; the point is that the soliton and
fermion fields are not relatively local and therefore cannot possibly be
in the same theory.

So far we have not mentioned one of the most important features of the
SGM and MTM --- their (quantum) integrability --- which
allows one to obtain exact results for various quantities in these theories.
In particular, the exact \sms\ of the two theories were first obtained within
the ``bootstrap approach''~\rZams\rKarThu,
and later using the quantum inverse
scattering method (for the MTM)~\rKore\ and by exploiting the quantum group
symmetry of the SGM~\rBeLe.
A subtle issue in all these approaches is how
to fix certain signs in the  scattering amplitudes of the charged particles.
This issue was never really resolved for the
SGM;
for the MTM it can be resolved by comparison with perturbation
theory, for example.

Even though these signs can not be ``detected''
 in scattering experiments in infinite volume, they
can       be observed in other
circumstances. The discussion of ``observables'' distinguishing the SGM and
MTM will be the general theme of sect.~4. In particular, in sect.~4.1 we
explain that the \sm\ signs are related to the statistics of the (fields
creating the) particles~\rKar\rZamtba. The fact that the solitons are bosons
implies that the sign of the SG \sm\ in the soliton sector is opposite to
that of the MTM in the fermion sector. As we have emphasized
previously~\rouriv\rourviii, \sm\ signs can be directly observed in the
finite-volume spectrum of a theory. In the remainder of sect.~4 we discuss the
finite-volume partition function of the SGM, mainly at $\beta^2=4\pi$,
where the corresponding MTM is free.
Starting with the exactly known partition function of the UV limit and using
some other input, we derive what we believe is the {\it exact} partition
function of the SGM at  $\beta^2=4\pi$. It provides an independent
confirmation of our claim about the bosonic nature of the solitons and the
signs of their scattering amplitudes.

In sect.~5 we discuss some extensions  of our observations.
One is about ``massive orbifolds'' of the SGM, defined as
perturbations of $c$=1 orbifold CFTs.
In particular, we propose exact \sms\ for one family of such theories.
 The other lies    outside the SGM/MTM context, and
concerns new classes of quantum field
theories, conformal and otherwise, which are ``essentially fermionic'' (\eg\ in
that their partition functions are only invariant under the subgroup of the
modular group generated by $S$ and $T^2$).
As an example, we present fermionic versions of the Virasoro minimal models.

Our conventions are collected in appendix~A, and some statistical mechanics
consequences of our results are discussed in appendix~B.

\medskip
In the previous pages we have summarized the
 essential    points of our paper.
The impatient reader
with some knowledge of the subject can just look up in the main text
whatever might have caught her/his attention. For the remaining potentially
disperate audience,
\eg~readers familiar with the SG
and MT models  but not conformal field theory, or {\it vice versa}, we have
provided a moderate amount of details in the following.
Readers who are not familiar with either subject are referred to~\rZams\ for
reviews of the former and~\rCarLH\rGinspLH\ for the latter subject.

\newsec{The UV CFTs}

\subsec{Gaussian CFTs}

The gaussian CFTs, of central charge $c=1$,
are constructed using a compactified free massless
scalar (or pseudo-scalar) field $\Phi(z,\bar{z})$ in two dimensions
(see appendix A~for our conventions).
By definition, $\Phi$ takes values on a circle,
\ie\ $\Phi \sim \Phi +2\pi r$, whose radius $r$
plays the role of a dimensionless coupling constant. Due to the decoupling of
left- and right-moving modes
(the field equation is $\partial\bar{\partial}\Phi(z,\bar{z})=0$), the theory
is actually ``enlarged''
to include two (almost) independent real fields $\phi(z)$ and
$\phib(\zb)$, with $\Phi(z,\zb)={1\o 2}(\phi(z)+\phib(\zb))$.
We will see later that both  $\phi(z)$ and $\phib(\zb)$ are compactified.

Because of IR divergences the fields $\phi(z)$ and $\phib(\zb)$ do not exist
as operators. They should be considered as ``mathematical building blocks''
of the gaussian models.
The actual operators are composites of derivatives and certain exponentials of
$\phi(z)$ and $\phib(\zb)$ (the exponentials must be well-defined with
respect to the compactification properties of
 $\phi$ and $\phib$).
The correlation functions of these operators can be
obtained from the formal 2-point functions of $\phi(z)$ and $\phib(\zb)$
(see appendix~A).  As will become clear in sects.~3 and~4, similar remarks
apply to the massive
field $\varphi$ ``underlying'' the SG and MT models.

To specify the gaussian CFTs we have to describe their Hilbert spaces, or
equivalently their        operator content.
By definition, these theories contain the holomorphic
and anti-holomorphic $U(1)$ currents $j(z)=i\partial \phi(z)$  and
$\bar{j}(\zb)=i\pb \phib(\zb)$. Consequently, the full Hilbert space
${\cal H}$  of a gaussian CFT can be written as
${\cal H}=\oplus_k {\cal H}_k \otimes \bar{{\cal H}}_k$,
where the ${\cal H}_k$ $(\bar{{\cal H}}_k)$ are $U(1)$-charged bosonic
Fock spaces in which the modes of $j(z)$ ($\bar{j}(\zb)$) act.
More explicitly (concentrating on the holomorphic side, the anti-holomorphic
side is treated in the same manner), expanding
$\phi(z) = q-i\alp_0 \ln z  + i\sum_{n\neq 0} {1\o n}\alp_n z^{-n}$,
the Fock space
${\cal H}_k$ is generated by repeatedly applying modes $\alp_{n<0}$
on the highest-weight state $|p_k\rangle$ ~($p_k\in \RR$)
satisfying $\alp_{n}|p_k\rangle
= \del_{n,0} p_k |p_k\rangle$ for $n\geq 0$. Hence
the Hilbert space ${\cal H}$ is completely specified
by the charges $(p_k,\bar{p}_k)$ labeling
the different sectors.

To find the
restrictions on the admissible     sets of $(p_k,\bar{p}_k)$,
it is simplest to consider
the operator product algebra~\rBPZ\ (OPA)
consisting of the fields that create the states in ${\cal H}$. The
$U(1)$$\times$$U(1)$ highest-weight state $|p\rangle \otimes |\bar{p}\rangle$
is created by the vertex operator
\eqn\Vmn{\eqalign{ V_{m,n}(z,\zb) & ~=~ :e^{ip\phi(z)+i\bar{p}\phib(\zb)}:
 ~=~ :e^{i({m\o 2r}+nr)\phi(z)+i({m\o 2r}-nr)\phib(\zb)}: \cr
 & ~=~ :e^{i {m\o r}\Phi(z,\zb) + 2 i n r \Phit(z,\zb)}: ~, \cr }}
where  $\Phit \equiv {1\o 2}(\phi-\phib)$
and
the quantum numbers $m=r(p+\bar{p})$ and $n=(p-\bar{p})/(2r)$
are referred to, respectively,
as ``momentum'' and ``winding''
(motivated by string theory)
or as ``electric'' and ``magnetic'' charges
(as in the Coulomb gas).
The normal ordering in~\Vmn\ is defined in appendix~A.

$V_{m,n}$ is a primary field~\rBPZ\ of conformal dimensions
$(\Delta_{m,n},\bar{\Delta}_{m,n})=({1\o 2}({m\o 2r}+nr)^2,
{1\o 2}({m\o 2r}-nr)^2)$,
so that its scaling dimension and (Lorentz) spin are
$d_{m,n}=\Delta_{m,n}+\bar{\Delta}_{m,n}=({m\o 2r})^2+(nr)^2$ and
$s_{m,n}=\Delta_{m,n}-\bar{\Delta}_{m,n}=mn$.
Other states in the charge sector $(p,\bar{p})$, alternatively
labeled by
$(m,n)$,
are created by the $U(1)$$\times$$U(1)$-descendants of $V_{m,n}(z,\zb)$.
These fields are generated by repeatedly taking the
operator product expansion (OPE) of the currents $j$ and
$\bar{j}$ with $V_{m,n}$; for instance
\eqn\OPEjV{ \eqalign{ j(w) V_{m,n}(z,\zb)~&=~
  {p\o w-z}~V_{m,n}(z,\zb)+
  :j(z)V_{m,n}(z,\zb):     +\ldots \cr
  ~&=~ {p\o w-z}~V_{m,n}(z,\zb)+p^{-1}\partial_z V_{m,n}(z,\zb)+\ldots~~,\cr}}
where the ellipsis stands for operators multiplied by $c$-functions that
are less singular as $w\to z$ than the terms shown.
The normal ordered product of operators in the above, and in similar equations
below, is defined by subtracting off the singular terms in the OPE of the
operators in question (\ie~the first line of~\OPEjV\ defines the normal
ordered product in this case).
Descendant fields also appear in the OPE of two vertex operators,
\eqn\OPEVV{ \eqalign{ V_{m,n}&(w,\wb)  V_{m',n'}(z,\zb) ~=~
 (w-z)^{pp'} (\wb-\zb)^{\bar{p}\bar{p}'}
   V_{m+m',n+n'}(z,\zb) + \ldots \cr   ~&=~
   (w-z)^{mn'+m'n} |w-z|^{2({m\o 2r}-nr)({m'\o 2r}-n'r)}
   V_{m+m',n+n'}(z,\zb) + \ldots~~,\cr}}
which
in a gaussian CFT is a special case of the exact equation
\eqn\OPEVVex{ :\! e^{i p \phi(w)}\! : ~:\! e^{i p' \phi(z)}\! : ~=~
  e^{-p p' \langle \phi(w) \phi(z)\rangle}
  :\! e^{i p \phi(w) + i p' \phi(z)}\!: }
and its ``complex conjugate''.

To complete the list of basic properties of the vertex operators
let us write down their $N$-point function:
\eqn\Vcorr{ \langle V_{m_1,n_1}(z_1,\zb_1) \ldots
   V_{m_\sN,n_\sN}(z_\sN,\zb_\sN) \rangle ~=~
   \del_{\Sigma m_i,0} \del_{\Sigma n_i,0}
  \prod_{i<j}^\sN
  (z_i-z_j)^{p_i p_j} (\zb_i-\zb_j)^{\bar{p}_i \bar{p}_j}~~.}

The OPA of a consistent CFT has to contain a single copy of the identity
operator, be closed and associative under the OPE, with each field
$A$ contain its conjugate field\foot{The conjugate $A^\ast$
is a field of the same conformal dimensions as $A$, satisfying
$A^\ast(w,\wb) A(z,\zb)$ ~=~ $(w-z)^{-2\Del_A} ~
  (\wb -\zb)^{-2\Delb_A} + \ldots$.}
$A^\ast$, and consist   of fields that are all {\it mutually local}~\rBPZ.
The latter is the requirement that
the $c$-function coefficients appearing in the OPE of any two operators
in the OPA are single-valued, so that correlators are single-valued as well.

We will now show that for CFTs with a $U(1)$$\times$$U(1)$ current algebra
these assumptions allow basically only two classes of OPAs. Defining
$L=\{ (m,n)\in \RR^2 ~|~V_{m,n}\in {\rm OPA} \}$, the above requirements
immediately imply that $L$ is an additive subgroup of $\RR^2$, satisfying
$(m,n),(m',n')\in L ~\Rightarrow ~ mn'+m'n\in \ZZ$, which follows from
mutual locality.
Note that this latter constraint
implies\foot{We ignore the less interesting
case of theories with only purely electric or magnetic charges. In these cases
the OPA is (a closed subalgebra of) that of an {\it un}compactified free
massless boson, obtained in
the limit $r\to \infty$ or $r\to 0$, respectively.}
 that $L$ is a {\it discrete}
subgroup of $\RR^2$,
and that the spin of any vertex
operator $V_{m,n}$ is   half-integer, $s_{m,n}= m n \in {1\o 2}\ZZ$.

Recall that  $\Phi={1\o 2}(\phi+\phib)$ lives on a circle of radius $r$,
so that $V_{m,0}$ is well-defined only if $m\in \ZZ$.
Assume, then, that $L$ contains some $(m,0)$ with $m\in \ZZ_{\neq 0}$
and another $(m',n')$ with $n'\neq 0$.
More precisely, let $m$ be the smallest positive integer such that
$(m,0)\in L$, and $n'$ the smallest positive real number such that
$(m'',n')\in L$ for some $m''\in \RR$. Then
there exists a unique $m'\in \RR$
such that $0<m' \leq m$ and $m'-m''\in m\ZZ$.
It follows that
$L=  \{ (m k +m' l, n' l) ~|~ k,l\in\ZZ\}$,
where $a\equiv m n'$ and $b\equiv 2m' n'$ must be positive integers.
We now use our freedom to rescale the radius $r\to \r r$
and at the same time transform
$L=\{ (m,n) \} \to \{ (m\rho,n/\rho) \}$, which has no effect on
the physics.\foot{Note that {\it a priori} $r$ is not a
physical parameter, since it enters the ``observable'' charges $(p,\bar{p})$
in combination with $(m,n)$. Only with the canonical lattices for the
$(m,n)$, given below in eq.~(2.6),
can $r$ be considered a meaningful parameter
characterizing a class of theories.}
For $b$ even we rescale $r \to n' r$ so that the lattice becomes
$L_{{\rm e}}(a,b)=\{(a k+{1\o 2}b l,l)~|~k,l\in\ZZ\}$. For $b$ odd,
$r \to 2 n' r$, and $L_{{\rm o}}(a,b)=\{(2a k+b l,{l \o 2})~|~k,l\in\ZZ\}$.
In both cases, the complete set of allowed lattices is labeled by
the integers $a$ and $b$ satisfying $0<b\leq 2a$.
Note that always $L_{\rm e}(a,b) \subseteq L_{\rm e}(1,2)$ and
$L_{\rm o}(a,b)\subseteq L_{\rm o}(1,1)$, so that for given $r$ there are
exactly
two ``maximal'' lattices of electric and magnetic charges leading to an
acceptable OPA of vertex operators, namely,
\eqn\Lbf{ \eqalign{ L_\rb~=~L_{\rm e}(1,2)~&=~\{ (m,n)~|~m,n\in \ZZ \}~~,\cr
  L_\rf ~=~L_{\rm o}(1,1)~&=~\{ (m,n)~|~m\in 2\ZZ,n\in \ZZ ~~{\rm or}~~
     m\in 2\ZZ+1,n\in \ZZ+{\textstyle {1\o 2}} \}~~.\cr} }

We emphasize that $L_\rb$ and $L_\rf$
are not equivalent; an obvious difference, that explains
our notation, is that the OPA corresponding to $L_{\rf}$ contains operators
with half-odd-integer spin,
whereas $L_\rb$ corresponds to only integer spin.
We will refer to the CFTs whose OPA is specified by $L_{\rb}$ and $L_{\rf}$
as the {\it bosonic} and {\it fermionic gaussian} CFTs, respectively.\foot{The
bosonic theories have been extensively discussed in the literature;
the fermionic ones were considered
by Friedan and Shenker~\rFS.}
Note that the OPAs of the two theories are generated by a quartet of
``fundamental''
operators, which can be chosen to be
$V_{0,\pm 1}$ and $V_{\pm 1,0}$ in
the bosonic case and $V_{\pm 1,\pm 1/2}$ in the fermionic case; these
quartets are of course {\it not} mutually local with respect to one another.

Concerning the theories based on proper sublattices of $L_\rb$ and $L_\rf$
we should say the following. Their correlation functions satisfy all the
standard axioms.
However, we will  see below (footnote~9)
that the partition functions of these theories are not
invariant under exchange of ``space'' and ``time''. In other words, the
euclidean covariant correlation functions of these models cannot be obtained
from a path integral with a euclidean invariant action.
This may sound strange, but just seems to show that a path integral is not
the only way to get a consistent set of correlation functions.
If for a {\it massive} theory, \eg~a perturbation of the above models,
 a set of Green's functions satisfying the
Wightman
axioms is sufficient to lead to a sensible particle interpretation
is quite a different question.
Be that as it may, the theories based on sublattices of $L_\rb$ and $L_\rf$
will play no role in the rest of this paper.

\medskip

We have to digress for a moment to discuss
 the statistics of the fields in the bosonic and fermionic gaussian models.
The commutation relations
$V_{m,n}(w,\wb) V_{m',n'}(z,\zb)=(-1)^{m n'+m' n} V_{m',n'}(z,\zb)
V_{m,n}(w,\wb)$ for $w\neq z$,
which follow from~\OPEVV,
are not the ``standard'' ones for generic $(m,n)\neq (m',n')$,
 neither in the bosonic nor the fermionic gaussian CFT.
To obtain the standard ones,
namely with ``fermionic'' operators (defined by $s\in \ZZ+{1\o 2}$)
anti-commuting among themselves and ``bosonic'' operators ($s\in \ZZ$)
commuting with all others, one has to multiply
the $V_{m,n}$ (and their descendants, of course) by appropriate ``Klein
factors''.    These are unitary operators that commute with all observables,
\ie~just multiply states by phases  that are constant
on sectors of given global charge.\foot{This implies, in particular,
that Klein factors cannot change the commutation relations of a field with
itself. And indeed, these  commutation relations
are already correct for the $V_{m,n}$.}
 In our case the global charges
are $(m,n)$, and the Klein-transformed vertex operators
$\hat{V}_{m,n} \equiv K_{m,n} V_{m,n}$ creating the charged sectors should
satisfy
\eqn\KVCR{ \hat{V}_{m,n}(w,\wb) \hat{V}_{m',n'}(z,\zb) ~=~
     (-1)^{2m n'} \hat{V}_{m',n'}(z,\zb) \hat{V}_{m,n}(w,\wb) ~~~~
             {\rm for}~~w\neq z }
in both the bosonic and the fermionic CFTs. The Klein operators $K_{m,n}$
are defined by
\eqn\Kdef{ K_{m,n}~|m_0,n_0\rangle ~=~ K_{m,n}(m_0,n_0)~|m_0,n_0\rangle ~~,}
and have to obey $K_{m,n}(m_1,n_1) K_{m,n}(m_2,n_2)= K_{m,n}(m_1+m_2,n_1+n_2)$.
Eq.~\KVCR\ then leads to
$K_{m,n}(m',n')/K_{m',n'}(m,n) =
(-1)^{m n' - m' n}$, which together with the other requirements on the
$K_{m,n}$ has as general solution
\eqn\genK{ K_{m,n}(m',n') ~=~ e^{ \pi i (\alp m n' +(\alp-1) m' n)} ~~~, ~~~~~
               \alp\in \RR  ~. }

\no Note that $K_{m,n} ~K_{m',n'} = K_{m+m',n+n'}$, ~$K_{0,0}=1$, and
$K_{m,n}~V_{m',n'} = K_{m,n}(m',n')~V_{m',n'}$ $K_{m,n}$.
For the conjugate of a vertex operator we have
 $\hat{V}^\ast_{m,n} = V^\ast_{m,n}~K^\ast_{m,n} = V_{-m,-n}~K_{-m,-n}
 = K^{-1}_{m,n}(m,n)~\hat{V}_{-m,-n}$.
The OPE of
Klein-transformed vertex operators is identical to~\OPEVV, except that the
rhs should be multiplied by   $K^{-1}_{m',n'}(m,n)$.

The choice of $\alpha$ does affect certain correlation functions and the
parity properties of the fields.
In a CFT this $\alp$-dependence would presumably not be considered
``observable''.  In a perturbed CFT, however, some correlation functions of
the perturbing field $V$ certainly are observable (they determine the
finite-volume energy levels, see sect.~4),
and one should choose
$\alpha$ so that $\hat{V} = V$. We will be interested in $V_{k,0}$ and
$V_{0,k}$ perturbations, and therefore choose $\alp=0$ in the first and
$\alp =1$ in the second case.

\medskip

Returning to the bosonic and fermionic gaussian models, it is illuminating
to reinterpret their operator content as follows.
Looking at the OPE of $\phi(w)$, $\phib(\wb)$ with
$V_{m,n}(z,\zb)$ and taking $w,\wb$ around $z,\zb$ one sees
that $V_{m,n}(z,\zb)$ creates a ``jump''
of size $2\pi n r$ in $\Phi$ at $(z,\zb)$,
and one of size $\pi m/r$ in $\Phit$. Therefore~\rFS\ one should identify
$(\Phi,\Phit) \sim (\Phi,\Phit)+(2\pi nr,\pi m/r)$ for   all $(m,n)\in
L_{\rb,\rf}$ for the bosonic/fermionic model. In other words, $(\Phi,\Phit)$
lives  on a torus,
\eqn\Lambf{ (\Phi,\Phit) \in \RR^2/\Lambda_{\rb,\rf}(r) ~, ~~~~~~
      \Lambda_{\rb,\rf}(r)~=~\{ (2\pi n r, \pi m /r)~|~
       (m,n)\in L_{\rb,\rf} \}~. }
Given these target spaces for $(\Phi,\Phit)$ in the bosonic/fermionic model,
their OPA consists exactly of all well-defined vertex operators
$V_{m,n} =  \exp({i m\o r}\Phi + 2i n r \Phit)$.

An interesting consequence of the above way of defining the bosonic
and fermionic gaussian CFTs is that it makes the observation of {\it duality},
\ie ~the equivalence of certain pairs of theories with different $r$,
clear from the start. Namely,
note that
\eqn\duality{ \Lambda_\rb\Bigl({1\o 2r}\Bigr)~=~\bigl(\Lambda_\rb(r)\bigr)^t
  ~~,~~~~~~~~
  \Lambda_\rf\Bigl({1\o r}\Bigr)~=~\bigl(\Lambda_\rf(r)\bigr)^t ~~,}
where
`$t$'
denotes reflection with respect to the
$\Phi=\Phit$ line.
This means that the bosonic (fermionic)
theory at ${1\o 2r}$ (${1\o r}$) can be obtained from that at $r$
by the simple field redefinition
$(\Phi,\Phit)\to (\Phit,\Phi)$,\foot{Note that
$\Phi$ and $\Phit$ have opposite parities under space (and time) reflection,
one  being a scalar, the other a pseudo-scalar. However, all correlation
functions are independent of the parity of $\Phi$, so the change of parity
accompanying a duality transformation is unobservable. As we will see later,
in a perturbed gaussian CFT the parity of $\Phi$ is not arbitrary, but
rather determined dynamically.}
 or equivalently $(\phi,\phib)\to (\phi,-\phib)$.
The effect of such
a field redefinition on the vertex operators is $V_{m,n} \to V_{n,m}
{}~(V_{2n,m/2})$ for the bosonic (fermionic) theory.
Note that the self-dual radii of the bosonic and fermionic theories are
different, being $r={1\o \sqrt{2}}$
in the bosonic case
(level one $SU(2)$-WZW model
with the unique modular invariant partition function)
and $r=1$ in the fermionic case (free Dirac point, see below).

Duality is more commonly demonstrated at the level of
partition functions.
The well-known calculation
of the torus partition function within the operator
formalism leads to
\eqn\pfbf{ \eqalign{ Z_{\rb,\rf}(q, r) ~&=~ {\rm Tr}_{{\cal H}_{\rb,\rf}(r)}
     q^{L_0-1/24}\qb^{\bar{L}_0-1/24}  \cr
   &=~ |\eta(q)|^{-2} \sum_{(m,n)\in L_{\rb,\rf}}
  q^{ {1\o 2}({m\o 2r}+nr)^2} \qb^{ {1\o 2}({m\o 2r}-nr)^2}~~,\cr} }
where $\eta(q)=q^{1/24}\prod_{k=1}^\infty (1-q^k)$ is the Dedekind eta
function, $q=e^{2\pi i\tau}$ ($\tau$ is the modulus of the torus), and
duality
\eqn\pfdual{ Z_{\rb}\Bigl(q, {1\o 2r}\Bigr)~=~Z_{\rb}(q, r)~~,~~~~~~~~
  Z_{\rf}\Bigl(q, {1\o r}\Bigr)~=~Z_{\rf}(q, r)~~}
is manifest.
Alternatively, $Z_{\rb}$ can be obtained
within the
euclidean path-integral formalism as the partition function of a compactified
free massless scalar   field with periodic boundary conditions
along the two cycles of the torus. This derivation  (which employs
zeta function regularization) makes clear the full modular invariance
of $Z_{\rb}(q, r)$,
namely $Z_{\rb}(q, r)=Z_{\rb}(e^{2\pi i}q, r)=Z_{\rb}(\tilde{q}, r)$,
where $\tilde{q}=e^{-2\pi i/\tau}$.\foot{Modular
invariance can of course be verified directly from \pfbf;
in particular, generalizing \pfbf\ and
denoting by $Z_L(q,r)$ the partition function corresponding
to an electric/magnetic charge lattice $L$, Poisson resummation
(see \eg\ \rGinspLH) leads to $Z_L(q,r)=Z_{\hat{L}}(\tilde{q},r)$
where $\hat{L}=\{ (\hat{m},\hat{n})\in \RR^2~|~m\hat{n}+n\hat{m}\in\ZZ ~\forall
(m,n)\in L \}$.  Hence $Z_L(q,r)=Z_L(\tilde{q},r)$ iff $L=\hat{L}$, which
holds for $L=L_{\rb,\rf}$ but not for any proper sublattices
of $L_{\rb,\rf}$. This last observation explains our earlier remark concerning
the theories based on such sublattices.
Note that for $c$=1 CFTs with a $U(1)$$\times$$U(1)$ symmetry, this result,
together with our classification of such theories given above,
extends those of~\rKir, where rationality of the CFT was assumed.}
$Z_{\rf}$, on the other hand, can easily be seen
not to be fully modular invariant but rather to satisfy   only
$Z_{\rf}(q, r)=Z_{\rf}(e^{4\pi i}q, r)=Z_{\rf}(\tilde{q}, r)$.
This invariance under
$\Gamma'$, the subgroup of the modular group $\Gamma$ generated by
$T^2$ and $S$ ($\Gamma$ itself is generated
by $T$: $\tau\to\tau+1$ and $S$: $\tau\to -1/\tau$) suggests that $Z_{\rf}$
corresponds to a euclidean  path integral with {\it anti}-periodic
boundary conditions along both cycles of the torus, appropriate for fermions.
This is is indeed the case~\rFP, as we will discuss in
subsect.~2.2.

\medskip
To conclude this subsection we
mention
the internal symmetries of the
gaussian CFTs that will be relevant later. For generic $r$ both the bosonic
and fermionic theories have an $O(2)$$\times$$O(2)$ symmetry, the symmetry
group of the toroidal target space in which $(\Phi,\Phit)$ lives.
We will
denote the two $O(2)$'s by $O(2)$ and $\Ot(2)$.
In a similar notation, they decompose into $\ZZ_2$$\times$$U(1)$ and
$\tilde{\ZZ}_2$$\times$$\tilde{U}(1)$. The $U(1)$, $\tilde{U}(1)$ act as shifts
on $\Phi$, $\Phit$,  while the $\ZZ_2$, $\tilde{\ZZ}_2$ are generated by
$R:~(\Phi,\Phit) \to (-\Phi,\Phit)$ and
$\tilde{R}:~(\Phi,\Phit) \to (\Phi,-\Phit)$, respectively.

\subsec{The Thirring model and its bosonization}

The (massless) Thirring model~\rThir\ is a (1+1)-dimensional QFT of a
massless Dirac fermion $\Psi(x)$ with a four-fermion self-interaction,
\eqn\TMlag{  {\cal L}_{{\rm TM}}
  ~=~ i\bar{\Psi} \rlap/\partial \Psi
                               + {\textstyle {\lam\o 2}} J_\mu J^\mu }
(here $\bar{\Psi}=\Psi^\dagger \gam^0$).
The (formal) field equation is
\eqn\TMeq{ i \rlap/\partial \Psi(x)~ =~ -\lam \rlap/J(x) \Psi(x)~~.}
To make this equation meaningful, in particular to give absolute meaning to
the dimensionless real coupling $\lam$, the normalization of $\Psi$ has to
be specified and the rhs, as well as
the current $J^\mu(x)$ itself (formally $\bar{\Psi}(x)\gam^\mu \Psi(x)$),
have to be carefully regularized.\foot{Our conventions below are
equivalent to those of
Johnson~\rJohn, which can be reproduced from the parametrization of~\rKlai\
by setting $g=\lam$ and $\sigma=-\lam/2$, or from that of~\rDAFZ\ by
taking $g=-\lam$ and $c=(\pi[1-({\lam\o 2\pi})^2])^{-1}$.
Coleman's $g$~\rCole\ is related to $\lam$ via
$\lam=-g/(1+{g\o 2\pi})$,
and Mandelstam's $g$~\rMand\ differs from Coleman's just by sign.}

The lagrangian \TMlag\ has an $O(2)$$\times$$O(2)$ symmetry, generated
by $\pmatrix{\Psi_1\cr \Psi_2\cr} \to \pmatrix{e^{i\th_1}\Psi_1\cr
e^{i\th_2}\Psi_2 \cr}$, $\th_{1,2}\in \RR$,
$\pmatrix{\Psi_1\cr \Psi_2\cr} \to \pmatrix{\Psi_2\cr \Psi_1 \cr}$, and
$\pmatrix{\Psi_1\cr \Psi_2\cr} \to \pmatrix{\Psi_1^\ast \cr \Psi_2^\ast \cr}$,
in the representation (A.1) of the Dirac matrices.
This symmetry also exists at the quantum level, as the exact correlation
functions show.
The         next thing to note is that the theories at
$\lam$ and  $-\lam$
are equivalent, as they are related by a simple
field redefinition
$\Psi=\pmatrix{\Psi_1\cr \Psi_2\cr} \to \pmatrix{\Psi_1\cr \Psi_2^\ast \cr}$
(having the effect of exchanging
$J^\mu$ and the axial current $\tilde{J}^{\mu}=\eps^{\mu\nu}J_\nu$).
Again, less  formally this {\it Thirring duality} can be seen in the exact
correlation functions
  (this was also noticed in~\rLuth).
These correlation functions also show that
the allowed region of the coupling is
$-2\pi < \lam < 2\pi$.

It is well known that the solution of the Thirring model can be
expressed
in terms of a free massless
(pseudo-)scalar field, namely the correlation functions
of $\Psi$ are identical to correlators of certain vertex operators of the
form~\Vmn. This is the so-called {\it bosonization} of the model, whose
history goes back (at least) to~\rGlas. Less known, apparently, is
the precise gaussian CFT that is equivalent to
the Thirring model at a given $\lam$.
Since the components of the Thirring field $\Psi$ are of Lorentz spin
$\pm {1\o 2}$ this gaussian CFT must be fermionic, the components corresponding
to $V_{\pm 1,\pm {1\o 2}}$. Up to $O(2)$$\times$$O(2)$ transformations
there are two inequivalent choices of bosonization, related by duality.
They are
\eqn\bostion{ \eqalign{ &({\rm i}) ~~~~~~~~~~
  \sqrt{2\pi} ~\Psi_1(x)  ~\leftrightarrow~
                                     \hat{V}_{1,{1\o 2}}(z,\zb)~~,~~~~~~~~~
           \sqrt{2\pi} ~\Psi_2(x) ~\leftrightarrow~
                                      \hat{V}_{1,-{1\o 2}}(z,\zb)~~,\cr
      &({\rm ii}) ~~~~~~~~~
  \sqrt{2\pi} ~\Psi_1(x)  ~\leftrightarrow~
                                       \hat{V}_{1,{1\o 2}}(z,\zb)~~,~~~~~~~~~
           \sqrt{2\pi} ~\Psi_2(x) ~\leftrightarrow~
                                       \hat{V}_{-1,{1\o 2}}(z,\zb)~~,\cr}  }
the factors of $\sqrt{2\pi}$ arising from the different but standard
normalizations of the kinetic term in Minkowski space QFT and euclidean
CFT.

Furthermore, the scaling dimension of $\Psi$
is $d_\Psi = {1\o 2}[1+({\lam\o 2\pi})^2]/[1-({\lam\o 2\pi})^2]$,
which can be read off from the 2-point function~\rJohn\
(see also~\rKlai\rDAFZ). Consequently, the
compactification radius in the corresponding fermionic gaussian CFT
is identified  as
\eqn\roflam{ r(\lam)~=~
     \sqrt{ {1\pm {\lam\o 2\pi}} \o {1\mp {\lam\o 2\pi}} }~~. }
The two possiblities, related by the Thirring duality
$\lam \leftrightarrow -\lam$, are equivalent in view of the duality of
the fermionic gaussian CFT $r\leftrightarrow 1/r$.\foot{The fact that
bosonization of the Thirring model with a given coupling $g$,
in Coleman's notation, leads to the dual (fermionic) gaussian CFTs at
$r=(1+{g\o \pi})^{\pm 1/2}$ was already noticed in~\rFP.
We emphasize that
it simply reflects the above-mentioned duality of
the Thirring models at $g$ and $-g/(1-{g\o \pi})$.}
We see that the self-dual $r=1$ fermionic gaussian CFT  is
the bosonized Thirring model at $\lam=0$,
\ie~the free massless Dirac
theory. Note that in this case the bosonized $\Psi_1$ ($\Psi_2$) is
of conformal weights $({1\o 2},0)$ ($(0,{1\o 2})$) according to \bostion,
hence holomorphic (anti-holomorphic),
in agreement with our conventions in appendix A.

We can also derive~\roflam\ directly in CFT, by writing down the CFT analog
of~\TMeq.  This will
explicitly exhibit the definition of the current $J_\mu$ we use,
and show  how
the two options in \roflam\ are correlated with those in \bostion.
We have (see \OPEjV)
\eqn\eomi{ \partial \hat{V}_{1,-{1\o 2}}(z,\zb) ~=~ {i \o 2}~({1\o r} -r)~
                         : \partial\phi(z)~\hat{V}_{1,-{1\o 2}}(z,\zb): ~~,}
and from~\OPEVVex\
and the properties of the Klein factors
\eqn\jope{\eqalign{ \hat{V}_{1,{1\o 2}}^\ast(w,\wb)~
      & \hat{V}_{1,{1\o 2}}(z,\zb)
  ~=~ |w-z|^{-{1\o 2}({1\o r}-r)^2}~(w-z)^{-1} \cr &\times ~\biggl[1-
  {i\o 2} ({1\o r}+r)(w-z)\partial\phi(z) -
  {i\o 2} ({1\o r}-r)(\wb-\zb)\pb\phib(\zb) + \ldots\biggr] ~. \cr }}
Now define
\eqn\jdef{ J(z) =
    \lim_{w\to z} A\biggl[ |w-z|^{{1\o 2}({1\o r}-r)^2}
       \hat{V}_{1,{1\o 2}}^\ast(w,\wb)
       \hat{V}_{1,{1\o 2}}(z,\zb)-{1\o w-z} \biggr]
      = -{i\o 2}({1\o r}+r)\partial\phi(z)~~,}
where $A$ denotes a suitable average,
\eg\ $A[\ldots] \equiv \oint_{z} {dw\o 2\pi i} {1\o w-z} \ldots$ ~(the contour
integral being around $z$),
which gets rid of the ``non-covariant'' term proportional to
${\wb-\zb \o w-z}$
in~\jope. The above,    together with      the analogous
equations for the fields of opposite chirality,
then give
\eqn\eomii{
 \eqalign{ \partial \hat{V}_{1,-{1\o 2}}(z,\zb) & ~=~ -{1-r^2\o 1+r^2}~:J(z)
  \hat{V}_{1,-{1\o 2}}(z,\zb): ~~, \cr
  ~~ \pb \hat{V}_{1,{1\o 2}}(z,\zb) & ~=~-{1-r^2\o 1+r^2}
   ~:\bar{J}(\zb) \hat{V}_{1,{1\o 2}}(z,\zb): ~.}}
This is equivalent to~\TMeq\ with choice~(i) in \bostion\ and the upper case
in \roflam,
$\lam=-2\pi \, {1-r^2\o 1+r^2}$,
with $(J(z),\bar{J}(\zb))\leftrightarrow 2\pi(J_0(x)\pm J_1(x))$. The dual
bosonization choice, (ii) of~\bostion, is treated similarly. It corresponds
to $\lam = 2\pi \, {1-r^2\o 1+r^2}$ and
$(J(z),\bar{J}(\zb))\leftrightarrow
 2\pi(\pm J_0(x) +  J_1(x))$.
[In both cases $(J(z),\bar{J}(\zb))$ are defined so that they equal
$-{i\o 2} ({1\o r} + r)(\partial \phi(z), \pb \phib(\zb))$.]

\medskip
As
alluded to
earlier, a direct euclidean path-integral
calculation~\rFP\ of the partition function of the
Thirring model, with anti-periodic boundary conditions on $\Psi$ along both
cycles of the torus, yields the result~\pfbf\ for $Z_{\rf}(q, r(\lam))$.
The bosonic partition function $Z_{\rb}(q, r(\lam))$ is obtained~\rFP\ by
summing up
all four possible boundary conditions, demonstrating that
the bosonic gaussian CFT is~\rThST\ a GSO-projected
Thirring model.
At the level of the OPA, the bosonic theory is obtained
from the Thirring model
by         a ``twist''~\rDFMS\ with respect to the total fermion number
$(-1)^F=(-1)^{F_1+F_2+\bar{F}_1+\bar{F}_2}$ ($=(-1)^{2n}$ in the
fermionic gaussian language) followed by a projection onto the $(-1)^F=1$
sector. In this framework the (bosonic) vertex operators
$V_{0,\pm 1}$, $V_{\pm 1,0}$
play the role of {\it spin fields}~\rsFQS\
for the Thirring fermion, in analogy with $\sigma$ of the Ising
CFT being the spin field for the free Majorana fermion.
It is also possible to construct the Thirring model by twisting the bosonic
gaussian CFT, or, in a more unified picture, to view both CFTs as two
different mutually-local projections
of one nonlocal         ``theory''
based on $L= \{ (m,n)~|~m\in \ZZ, n\in {1\o 2}\ZZ \}$.

It is amusing to note the interplay between duality and the
above $\ZZ_2$ twist.
For instance, the equivalent bosonic theories at $r=1$ and $r={1\o 2}$
can be obtained by
twisting
either the free Thirring (self-dual)
point $\lam=0$, or of the interacting Thirring model at $\lam=\pm {6\pi\o 5}$.
There is no contradiction, though, as we would like to think of the
resulting bosonic theory as interacting anyhow
(cf.~sect.~4),
in the sense that it
contains a nontrivial interacting sector --- the $m\in 2\ZZ+1$ sector in the
$r=1$ representation, or $n\in 2\ZZ+1$ in the dual $r={1\o 2}$ one.
When twisting the free Thirring model this sector arises as the twisted
sector that survives the $(-1)^F=1$ projection, the ``non-interacting''
sector of the bosonic theory coming from the untwisted sector of the
free Thirring model. When twisting the
$\lam=\pm {6\pi\o 5}$
Thirring model the situation is more intricate; the interacting and
non-interacting sectors of the bosonic theory are built of
$(-1)^F=1$ operators in both the twisted and untwisted Thirring sectors.

\newsec{Identification of the Sine-Gordon and Massive Thirring Models}

We are now ready to discuss the massive QFTs whose UV limits are the bosonic
and fermionic gaussian CFTs described in the previous section. It has proven
very fruitful, both from a conceptual as well as a practical point of view
(see \eg~\rCole\rZamflow\rLudCar\rTanig), to view such theories as relevant
perturbations of their UV CFTs.

We will consider only perturbations by a single relevant ($d$$<$$2$), spinless
($s$=$0$), real operator, \ie~either by $V_{m,0}^{(\pm)}$,
$m$$=$$1,2,\ldots <2\sqrt{2} r$
(with $m$ only even in the fermionic case),
 or $V_{0,n}^{(\pm)}$, $n$$=$$1,2,\ldots <
\sqrt{2}/r$.
[Here for any operator $A$, $A^{(+)} \equiv {1\o\sqrt{2}}
(A+A^\ast)$ and $A^{(-)}\equiv {1\o \sqrt{2} i}(A-A^\ast)$, where the conjugate
operator $A^\ast$ was defined in sect.~2.1.]
Due to the $O(2)$$\times$$O(2)$ symmetry of the unperturbed theory,
perturbations
by any linear combination of the purely electric operators $V_{m,0}^{(+)}$
and $V_{m,0}^{(-)}$ lead to the same massive theory, and similarly for the
magnetic operators  $V_{0,n}^{(+)}$ and  $V_{0,n}^{(-)}$. We can therefore
restrict attention to the  $V_{m,0}^{(+)}$ and  $V_{0,n}^{(+)}$ perturbations
with {\it positive} couplings.
Using the ``counting argument''~\rTanig\ it is not difficult to see
that any such perturbation is integrable (see \eg~\rHenSal).

The euclidean action of the perturbed theories can therefore be written as
\eqn\paction{ A_{\rb,\rf}(r,V) ~=~ A_{\rb,\rf}(r) ~+~ \mu
   \int        V ~~,}
where $A_{\rb,\rf}(r)$ is the action of the bosonic/fermionic gaussian CFT
with $\Phi$ living on a circle of radius $r$,
$V$ is either $V_{m,0}^{(+)}=\sqrt{2}\cos (m\Phi/r)$ or
$V_{0,n}^{(+)}=\sqrt{2}\cos(2nr\Phit)$, and $\mu>0$ is of mass dimension
$y =2-d_V>0$. By duality the theories described by $A_{\rb}(r,V_{m,0}^{(+)})$
and $A_{\rb}({1\o 2r},V_{0,m}^{(+)})$ are identical, and the same is true for
$A_{\rf}(r,V_{0,n}^{(+)})$ and $A_{\rf}({1\o r},V_{2n,0}^{(+)})$.
Except for these identifications, the theories defined above are distinct.

Since both $\Phi$ and $\Phit$ are free fields in the massless theory, it is
clear that       ``underlying''  all of the massive theories~\paction\
is a field
$\varphi$ ($\propto \Phi$ or $\Phit$) that obeys a SG equation.
This
demonstrates one of the weaknesses of a lagrangian approach on the
non-perturbative level, where often
there is no direct relation between the particle
spectrum and the fields in the lagrangian: Distinct QFTs can
``contain'' fields
that obey the same equations of motion (even when all renormalization effects
are taken into account). In other words, a lagrangian does not in general
define a unique QFT.\foot{This is a statement on the quantum level; further
non-uniqueness might be associated with passing from the classical to the
quantum lagrangian.}

The best one can hope for in a lagrangian approach is that the equations of
motion for the ``fundamental fields'' of the theory follow from a lagrangian.
(By fundamental fields we mean operators whose derivatives and OPEs generate
the full operator algebra
of the theory.) As we will discuss below, this is the
case for the MTM where
the components of $\Psi$ and $\Psi^\dagger$,~\ie~the
massive analogs of $V_{\pm 1, \pm {1\o 2}}$,
are the fundamental fields.

The perturbations of the bosonic gaussian CFTs, on the other hand, illustrate
that the fundamental fields do not necessarily obey lagrangian equations of
motion. Here the fundamental fields
are presumably the
massive analogs of
$V_{\pm 1,0}$ and $V_{0,\pm 1}$. But in the UV limit these fields obey first
order equations of motion (derived similarly to~\eomii) and there is no first
order Lorentz-invariant lagrangian for fields of Lorentz spin~$0$. Since there
is already no lagrangian for the fundamental fields in the massless limit,
there is no hope for one describing  the fundamental fields in the  theory
perturbed by a relevant operator.

Sticking to a lagrangian approach, in cases like the perturbed bosonic CFTs
one must be content
with a lagrangian for a non-fundamental field, \eg~the
SG field $\varphi$.
To discover the full operator content of the theory one should look for soliton
solutions, first on the classical level, and then try to construct the
corresponding super-selection sectors on the quantum level. This is not easy,
in general, and it is one of the many
advantages of describing the massive theories by~\paction\   that the
super-selection sectors are
manifest --- each super-selection sector corresponds to the
union of sectors of the CFT that have the same global charge with respect
to the symmetry that remains unbroken in the perturbed theory.

Let us now
 complete the lagrangian description of the
theories specified by~\paction, starting with the bosonic theories.
This equation and the form of the allowed $V$ implies that in
terms of the SG lagrangian~\lagSG,
\eqn\pbvsSG{\eqalign{ A_\rb(r,V_{k,0}^{(+)}) & ~\leftrightarrow~
 {\cal L}_{{\rm SG}}(\beta={\sqrt{\pi} k/r})~, ~~~~~~ \,
 \varphi\equiv {\Phi \o \sqrt{\pi}} ~\sim~ \varphi + {2\pi\o \beta} k \cr
  A_\rb(r,V_{0,k}^{(+)}) & ~\leftrightarrow~
 {\cal L}_{{\rm SG}}(\beta=\sqrt{   4 \pi} k r)~, ~~~~~~
 \varphi\equiv {\Phit \o \sqrt{\pi}} ~\sim~ \varphi +{2\pi\o \beta} k ~, \cr }}
where we used the fact that the SG coupling does not renormalize.
In contrast, the mass parameter $\alp_0$ in~\lagSG\ does;  its relation to
$\mu$ in~\paction\ therefore depends on the scale at which $\cos\beta\varphi$
is normal ordered~\rCole, and is irrelevant here.
Much more interesting is the dimensionless coefficient
$\kappa=\mu m_1^{-y}$ relating
the ``bare coupling''
$\mu$ to a {\it physical} mass scale $m_1$,
\eg~the mass of the lightest particle in the perturbed theory.
(In sect.~4.3 we will find the exact value of $\kappa$ in
a special case.)

Note that for the $V_{k,0}^{(+)}$ and $V_{0,k}^{(+)}$ perturbations
in~\pbvsSG\       exactly      $k$
periods of the cosine potential fit on the circle on which $\varphi$ lives.
These theories were denoted by SG$(\beta,k)$ in~\rourviii. Intuitively
one expects that  SG$(\beta,k)$ will have a $k$-fold degenerate vacuum, and
that for $k\! >\! 1$ these QFTs will be ``kink theories'', \ie~have nontrivial
restrictions on their multi-particle Hilbert spaces.
For example,
in finite volume with periodic boundary conditions
 the only allowed soliton states in  SG$(\beta,k)$ are such that
(\#~solitons~$-$~\#~antisolitons)~$\in k\ZZ$.
We will discuss these theories in more detail in~\rournext.

The ordinary SG model corresponds to the $k$=1 case, \ie~we identify
SG$(\beta)$ = SG$(\beta,1)$ = $A_\rb(r\! = \!{\sqrt{\pi}/\beta},V_{1,0}^{(+)})=
 A_\rb(r\! =\! {\beta/\sqrt{  4\pi}},V_{0,1}^{(+)})$.
This is in
accordance
with the
common belief that the UV limit of the SG models are bosonic gaussian CFTs.
As in other cases, arguments for this would presumably be given in the
context of lattice models,
either by a study of partition functions in the thermodynamic limit, or by
exhibiting a field which obeys the SG equation in the continuum limit.
Neither of these methods allows one to see the difference between the
bosonic and fermionic gaussian models. Knowledge of the finite-temperature
infinite-volume partition function is equivalent to that of the
finite-volume
ground state energy, which is identical
in SGM and MTM (cf.~sect.~4), and a field obeying the SG equation can also be
written down for perturbed fermionic gaussian CFTs.
One might therefore in fact have some doubts about our above identification
and wonder if the SGM is perhaps a perturbed  {\it fermionic} CFT.

However, there is a simple argument to dispel such doubts, based on the
periodicity of the field $\varphi$.
Namely, exactly as in~\pbvsSG\ one sees
that the theories   $A_\rf(r\! = \! {\sqrt{\pi}k/\beta},V_{k,0}^{(+)})=
                     A_\rf(r\! = \! {\beta/\sqrt{\pi} k},V_{0,{k\o 2}}^{(+)})$,
$k\in 2\NN$, involve an  ``underlying'' field $\varphi$ obeying the SG
equation with coupling $\beta$; but now
the number       $k$       of periods of the
potential that fit on the ``target space'' of $\varphi$,
                 cf.~\Lambf,
is always {\it even}.\foot{Elaborating on an earlier
remark in this section, we now see that distinct QFTs can involve
a field obeying the same equations of motion even when the periodicity of this
field is taken into account.}
Therefore none of these theories can correspond to the standard SGM, where
exactly one period fits on the target space
of $\varphi$.
We should also add that for the above perturbed fermionic CFTs $k/2$, not
$k$, is the degeneracy of the vacuum, as we will discuss in~\rournext.
(In the special case $k=2$ this just corresponds to the fact that the
vacuum of the MTM is non-degenerate, see below.)
 This is possible, and  not
in contradiction to our above remarks about  SG$(\beta,k)$, because the
SG ``field'' $\varphi$
is compactified and therefore not really a well-defined field,
neither in the massless nor the perturbed gaussian models;
 its periodicity just
encodes the operator content of the theory    (at least partially).
The conventional
claim that $\varphi$ creates the lowest bound state in SGM should be replaced
by the statement that $\sin\beta\varphi$ creates this particle,
cf.~sect.~4.4.

\medskip
We can now identify the soliton creation operators in the SGM. Note that for
a $V_{0,k}^{(+)}$ ($V_{k,0}^{(+)}$) perturbation of a bosonic
gaussian CFT the electric charge $m$ (magnetic charge $n$) is
super-selected also in the massive theory. We identify $m$ ($n$) as the soliton
number (=~\#~solitons~$-$~\#~antisolitons). This is confirmed by the following
fact, considering for example a
$V_{0,k}^{(+)}$ perturbation: $V_{m,0}$ creates a
discontinuity of $\pi m/r$ in $\Phit$ (cf.~sect.~2.1), corresponding
to a jump of ${\pi m\o \sqrt{\pi} r} = {2\pi m\o \beta}$ in the SG field
$\varphi =\Phit/\sqrt{\pi}$, in accord with the     well-known
normalization of
the topological charge in the SG model, ${\beta \o 2\pi}[\varphi(\infty)-
\varphi(-\infty)] \in \ZZ$.
The massive theory will provide a length scale over which the
discontinuity created by $V_{m,0}$
is smeared out.
The soliton/antisoliton creation operators are presumably the simplest
operators of soliton number $\pm 1$, namely massive analogs of $V_{\pm 1, 0}$.
Note that these operators have Lorentz-spin~0, \ie~are bosonic.
The work of Mandelstam~\rMand\ essentially shows that the above considerations,
phrased in the language of CFT, also hold in the massive theory.

\bigskip
Let us now     discuss the perturbed fermionic theories
$A_\rf(r,V_{2,0}^{(+)}), ~A_\rf(r,V_{0,1}^{(+)})$
in more detail.
Here we should be careful and use the Klein-transformed vertex
operators $\hat{V}_{m,n}$
throughout.          Using~\OPEVV\ and the properties of the Klein factors
we can then express the
perturbing field in terms of the fundamental fields
$\hat{V}_{\pm 1, \pm {1\o 2}}$ as
\eqn\fpert{ \eqalign{
   {1\o \sqrt{2}} \lim_{w\to z}
    |w-z|^{\mp {1\o 2}(r^2-r^{-2})}
    ~[\hat{V}_{1,{1\o 2}}^\ast(w,\wb) \hat{V}_{\pm 1,\mp{1\o 2}}(z,\zb)& +
   \hat{V}_{\pm 1,\mp{1\o 2}}^\ast(w,\wb) \hat{V}_{1,{1\o 2}}(z,\zb)] \cr
     ~~~ &=~ \cases{\hat{V}_{0,1}^{(+)}(z,\zb) \cr
                    \hat{V}_{2,0}^{(+)}(z,\zb) ~,\cr}
    \cr} }
where we take $\alp=1~(0)$ in \genK\ for the upper (lower) case.
In terms of the fermion of the TM we therefore have
\eqn\pTM{
   {1\o \sqrt{2}} ~ \lim_{y\to x}~|y-x|^{\mp {1\o 2}(r^2-r^{-2})}
        ~2\pi~\bar{\Psi}(y) \Psi(x)~~ \leftrightarrow~~
 \cases{  \hat{V}_{0,1}^{(+)}(z,\zb) \cr
          \hat{V}_{2,0}^{(+)}(z,\zb) ~,\cr} }
the upper/lower case corresponding to the bosonization choice (i)/(ii)
in \bostion.
The perturbations in question therefore really correspond to the addition
of a properly regularized~\rDAFZ\rCole\
mass term to the massless Thirring lagrangian.

To recapitulate, we identify $A_\rf(r,V_{0,1}^{(+)})$ as the MTM
at $\lam=-2\pi \, {1-r^2 \o 1+r^2}$ ~(or $g=\pi(1-r^2)/r^2$ in Coleman's
conventions), and $A_\rf(r,V_{2,0}^{(+)})$ as the MTM
at $\lam=2\pi \, {1-r^2 \o 1+r^2}$ ~($g=-\pi(1-r^2)$). The condition for
the relevance of the perturbation, $r\! < \! \sqrt{2}$ and
$r\! > \! 1/\sqrt{2}$,
respectively, translates into $-2\pi<\lam<2\pi/3$
{}~ ($g>-\pi/2$)
in both cases.
Note that within the {\it massive} TM there is {\it no}
duality between the attractive
($-2\pi<\lam<0$) and repulsive ($0<\lam<2\pi/3$) regimes.

The perturbations by $V_{0,{k\o 2}}^{(+)}$ and $V_{k,0}^{(+)}$ with
even    $k>2$  correspond to adding
suitably regularized terms of $k$-th order in $\Psi$ to the massless Thirring
lagrangian. They describe
``fermionic kink theories'',  and will be discussed in~\rournext.
Let us here just note the obvious fact that
the conserved global charge, $m$ ($2n$) for a $V_{0,{k\o 2}}^{(+)}$
($V_{k,0}^{(+)}$) perturbation of a fermionic CFT, is
fermion number, ~\#~fermions~$-$~\#~anti\-fermions.

\bigskip
Finally, we should elaborate on our remarks in the introduction concerning
Mandelstam's results~\rMand. He works directly with the massive theories,
where things are somewhat more involved than in the massless case, but we can
illustrate the basic points of his analysis by translating it into CFT
language. The starting point is to find operators that create a
``jump'' of one period
in the SG field $\varphi$, as candidates for soliton creation operators.
Mandelstam notices that such operators can either commute
or anti-commute, corresponding to, in our notation, massive analogs of
$V_{\pm 1,n}$ with $n\in \ZZ$ or $n\in \ZZ+{1\o 2}$, respectively (we choose
$\varphi\propto \Phit$ in this discussion). He pursues the study of the
simplest anti-commuting operators, $V_{\pm 1,\pm {1\o 2}}$, leading to the
bosonized MTM. His analysis works just as well for $V_{\pm 1,0}$, which leads
to the SGM.

The massive $V_{m,n}$ are identical in structure to the CFT ones; the only
differences are: (i) Instead of a compactified free massless field one uses a
compactified massive field $\varphi$ obeying the SG equation
to construct
 the $V_{m,n}$. ~(ii) Since for a massive field the left- and
right-moving modes do not decouple, one should express the ``dual'' of
$\varphi$ as a nonlocal functional of $\varphi$, \ie~use the analog of the
CFT expression
\eqn\Vnonloc{V_{m,n}(z,\zb) ~=~ :\exp\Bigl({{i m\o r}
  \int_{\infty}^{(z,\zb)} (dw\partial_w \Phit - d\wb \partial_{\wb} \Phit)
    ~+~ 2 i n r ~\Phit(z,\zb)}\Bigr):   ~~,}
ignoring Klein factors and other subtleties.
As emphasized previously, the
operators $V_{\pm 1,0}$ and $V_{\pm 1,\pm {1\o 2}}$ cannot belong to the same
theory since they are not mutually local.
In the perturbed theory
mutual locality of arbitrary vertex operators is
equivalent to requiring
that the ``Mandelstam strings'' in~\Vnonloc\
be
 ``invisible'';
the constraint $mn'+m'n\in \ZZ$
(see sect.~2)
can then be interpreted as a Dirac-like
quantization condition on the allowed electric and magnetic charges.

\newsec{Observables in SGM and MTM}

\subsec{$S$-Matrices}

In sect.~3 we have identified the SG and MT models as perturbed CFTs, which
immediately gave us a detailed understanding of their UV limits, in particular
how they differ. The next step is to
uncover
the difference between the massive theories.
Given
certain smoothness assumptions about the UV limit, to be discussed in
subsect.~4.2, we right away know that some correlation functions differ and
some are identical in corresponding SG and MT models. Namely, correlators
of fields whose UV limit is in one of the ``even'' sectors
$(m,n)\in 2\ZZ$$\times$$\ZZ$
will be identical, whereas
all other correlation functions will be distinct, since they contain some
fields from the ``odd'' sectors where the UV limits of SGM and MTM have
no fields in common.

It is of course
very difficult
(see however~\rSmirBook) to calculate these
correlation functions exactly, and in any case we are more interested in
directly ``observable'' quantities, \eg~\smes\ and finite-volume energy levels.
Considering the
SG/MT models          from now on as $V_{0,1}^{(+)}$-perturbed
CFTs, we know that the electric charge $m$ corresponds to soliton/fermion
number.
[In the notation of subsect.~2.1, the $O(2)$ symmetry of the CFT remains
unbroken; the parity $(-1)^m$ of the sectors in the massive theory refers
then to the $\ZZ_2$ subgroup of the unbroken $U(1)$.]
The bound states of the
solitons, respectively, fermions correspond to $m=0$,
\ie~one of the even sectors where the correlation
functions in SGM and MTM are identical. Therefore the \scamps\ involving only
bound states will be identical in both theories.

The \sm\ in the soliton/fermion sector is determined almost uniquely by the
general requirements of factorizable \smt\ for
an $O(2)$-doublet of scattering
particles~\rZams\rKTTW. The only ambiguity are CDD factors.
They have to be the same in SGM and MTM, up to a sign perhaps, to guarantee
that the bound state amplitudes, which can be obtained using the
bootstrap~\rZams\rKarThu, agree.
If one makes the very plausible assumption
that there are no bound states in the repulsive regime of the MTM,
the CDD factors must be trivial (=$\pm 1$)~\rKTTW,
and
             the only ambiguity that remains
is one overall sign
for the \sm\ in the soliton/fermion sector.

For the MTM the \sm\ so obtained has been checked perturbatively to third
order~\rWeisz, confirming in particular the choice of sign made
in~\rZams\rKTTW.
However, for the SGM there is no lagrangian for the soliton
fields which would allow for a perturbative check, and we in fact claim that
the overall sign of the SG \sm\ in the soliton sector is opposite to that of
the MTM!

There is a very simple argument for this: Let $S_{aa}^{aa}(\th)$ denote the
\sme\ for the scattering of two particles of species $a$ in a factorizable
\smt; $\th$ is the relative rapidity and we are assuming that the scattering is
purely elastic,
{\it i.e.}\foot{$S_{ab}^{cd}(\th_1-\th_2)$
denotes the amplitude for the process
 ~$a(\th_1)+b(\th_2)\to c(\th_2)+d(\th_1)$.}
 ~$S_{aa}^{cd}(\th)=\del_a^c \del_a^d S_{aa}^{aa}(\th)$
(as for the solitons in SGM and the fermions in MTM,
by charge conservation).
By considering
multi-particle states one can then easily show~\rZamtba\
that an {\it exclusion principle}
(in rapidity space)
holds  only if $S_{aa}^{aa}(0)=-(-1)^{F_a}
=\mp 1$ if particle $a$ is a boson/fermion.\foot{Ignoring the special case of
kink theories~\rourviii,  the only QFT we know for which $S_{aa}^{aa}(0)\neq
-(-1)^{F_a}$ are free bosons, which indeed
do
not satisfy an exclusion
principle.} Since presumably any interacting
(1+1)-dimensional
 QFT
satisfies an exclusion
principle, our identification of the SG solitons as bosons then proves our
claim.

A similar argument can be given using the relation between the parity of a
bound state and the sign of the residue of a pole in an appropriate channel
of the \sm. Such a relation follows from the following two statements:

\no {\bf (i)}~ Poles of the \sm\ corresponding to bound states with symmetric
(anti-symmetric) wave functions have positive (negative) imaginary residue,
when the \sm\ is considered as a function of the relative rapidity $\th$.

\no {\bf (ii)}~ Boson-antiboson ($b\bar{b}$) bound states with symmetric
(anti-symmetric) wave function have positive (negative) $C$ and $P$ parity.
For fermion-antifermion ($f\bar{f}$)  bound states this relationship is
reversed.

Statement~(ii)~is basically obvious for the corresponding unbound 2-particle
states (see~\eg~chapter~15 of~\rBjDr), and can therefore be expected to be
true at least for weak bound states. (i)~can be proved in potential
scattering and is also believed to hold in QFT~\rKarThu\rKar. By combining~(i)
and~(ii) we see that the relation between the $C$ and $P$ parity and the sign
of the residue will be opposite for the case of a $b\bar{b}$ and $f\bar{f}$
bound state. If we can establish that the parities of the bound states in SGM
and MTM are the same, our claim about the sign of the soliton \sm\ follows.
But it is well known how to determine the parities of the bound
states~\rZams\rKarThu. In a 2-particle basis of $C\! =\! \pm 1$ eigenstates,
$|s(\th_1)\bar{s}(\th_2)\rangle\pm |\bar{s}(\th_1)s(\th_2)\rangle$ for SGM,
say, the soliton-antisoliton ($s\bar{s}$) \sm\
is diagonal with amplitudes
$S_\pm(\th)$~ \rZams\rKarThu,
respectively. One finds that a pole corresponding to the $n$-th
bound state\foot{Recall that the number of bound states in SGM (and  the
corresponding MTM) is equal to the largest integer smaller than
$\lam\equiv{8\pi\o \beta^2}-1$; their masses are $m_n(\lam)=  2\, m \,
\sin{n\pi\o 2\lam}$, $n=1,2,\ldots <\lam$ in terms of the soliton (fermion)
mass $m$.}
appears in $S_+$ only for even $n$, in $S_-$ only for odd $n$ (and this fact
is obviously independent of the
overall
sign of the $s\sb$ \sm).

For completeness, we note that amplitudes involving one soliton/fermion and
a bound state have the same sign in SGM/MTM. This can be shown using the
bootstrap.

\medskip
So as to not always argue about the signs of \smes, we will in the
following subsections look at a rather different kind of observable that
distinguishes the SG and MT models, namely
the finite-volume spectrum, as encoded in the partition function.
The finite-volume partition function
has proven to be a useful and illuminating probe of
QFTs defined as perturbed CFTs. The reason is that it
provides an interpolation between the UV CFT for small volume, which
is usually understood completely, and the IR behaviour of the theory for
large volume, which allows one to obtain some information about the \sm, in
particular signs.
The arguments we use in this analysis are unrelated to those employed above,
and will, in particular, provide independent evidence for the claim that
the solitons are bosons.
The small-volume behaviour of a   partition function
can be determined in CPT, to which we now turn.

\medskip
\subsec{Conformal Perturbation Theory}

The idea of defining a massive
(1+1)-dimensional QFT as a (spinless) relevant
perturbation of a CFT has been around for a long time.
It
has become
particularly fruitful in the last few years, after many classes of CFTs
were understood in great detail.
In the case of integrable perturbations, this approach has led to
exact non-perturbative results when used in conjunction with other techniques,
like the bootstrap,
exploiting quantum group symmetries,
the thermodynamic Bethe Ansatz,
and,      more
generally, the numerical and analytical study of the finite-volume spectrum
of a non-scale-invariant  QFT.
(For all of this
see~\rTanig\rBeLe\rZamtba\rourii\rLMC\rouriii\rouriv\rFenExc\rFenInt\rourviii\
and references therein.)

The reason that makes this idea so powerful is that a relevant
(\ie~super-renorm\-al\-iz\-able)
perturbation
of a CFT is a very ``benign'' perturbation. For example, there are good
reasons~\rKMS\rouriii\
 to believe that a perturbative expansion around the UV CFT will have
a finite radius of convergence with appropriate IR and, if necessary (see
below), UV cutoffs.
Furthermore, the super-selection structure of the massive theory
is manifest, since that of the CFT is usually known, and the massive sectors
will simply be labeled by the charges that are conserved by the perturbation.
Of course, the structure within a sector, \eg~if there are bound states in the
vacuum sector, is not obvious in this approach either.
Still, one can
often ascertain facts which are not at all trivial in a standard lagrangian
approach; for example, as mentioned in subsect.~4.1, that the properties of
the bound states of the solitons/fermions in
the SGM/MTM are absolutely identical.

The expected ``smoothness'' of relevant perturbations of CFTs
is reflected in the finite-volume spectrum of the perturbed theory.
(We will always put the theory on a cylinder, \ie~use periodic boundary
conditions for bosons and anti-periodic ones for fermions.) The finite-volume
energy levels must be smooth functions of the ``volume'' of space,
namely the circumference $L$ of the cylinder.
For small volume the eigenstates
are simply labeled by the states of the UV CFT, and the $i$-th energy gap
$\hat{E}_i(L) = E_i(L)-E_0(L)$
above the ground state
behaves like~\rFSSC\
(for a perturbed {\it unitary} CFT at least)
\eqn\EsmallL{\hat{E}_i(L) ~\to~ {2\pi \o L} ~d_i ~~~~~~~~~{\rm as}~~L\to 0~,}
where $d_i$ is the scaling dimension of the UV conformal state
$|i\rangle$ created by $\phi_i$. The ground state energy itself behaves like
$E_0(L)\to -\pi c/6L$, where $c$ is the central charge of the (unitary) UV
CFT.
[In the cases considered the ground state level presumably does not cross
any other level for all $L\geq 0$, in other words, there is no
phase transition for nonzero temperature.]
The total momentum $P$ of a state is quantized in finite volume, and
we have the {\it exact} relation ~$P_i(L)={2\pi \o L} s_i$~ to the spin $s_i$
of the UV field $\phi_i$.   The finite-volume spectrum then provides a smooth
interpolation between the states of the UV CFT and those of the massive theory
in the IR, where
                               any
$\hat{E}_i(L)$ simply
approaches some sum        of masses.

To emphasize that defining a QFT as a perturbed CFT is
essentially a {\it non}-perturbative
definition of the theory, we remark that using the
``truncated conformal space approach'' of Yurov and Zamolodchikov~\rYZ\rLasMus\
one can in principle calculate the finite-volume energy gaps of the perturbed
theory to arbitrary accuracy; the method is non-perturbative, though numerical.
                         We will not use this technique here,
but instead consider the analytical calculation of the small-volume expansion
coefficients for the energy levels, \ie~CPT. The action of the perturbed CFT
is
\eqn\pactionii{ A_{\lam} ~=~ A_{{\rm CFT}} ~+~ \lam \int d^2\xi ~V(\xi) ~~,}
where the integral is over the cylinder,
$\lam=\kappa m^y$ for some dimensionless constant $\kappa$, $d=2-y$ is
the scaling dimension of $V$, and $m$ a mass scale in the perturbed theory,
the mass of the lightest particle in the cases we will consider.
The dimensionless gap scaling
functions $\hat{e}_i(\r)$ can then be expanded as
\eqn\eih{\hat{e}_i(\r)~\equiv~{L\o 2\pi} \hat{E}_i(L) ~=~ d_i
 ~+~ \sum_{n=1}^\infty ~\hat{a}_{i,n} ~{\r}^{ y n},
      ~~~~~~~\r ~\equiv~ Lm
                           ~,}
where
$\hat{a}_{i,n} \equiv a_{i,n} - a_{0,n}$  with
\eqn\ain{ a_{i,n} ~=~  -{(2\pi)^{1-y}~(- \kappa)^n \o n!~\langle i|i\rangle}~
    \int \prod_{j=1}^{n-1} {d^2z_j \o (2\pi|z_j|)^{y}}~
     \langle i| V(1,1)
    \prod_{j=1}^{n-1} V(z_j,\bar{z}_j)| i\rangle_{{\rm conn}} ~~.}
The correlators here are connected
(with respect to the ``in- and out-states'' created by $\phi_i$,
$\phi_0$ being the identity operator)
critical $(n+2)$-point functions  on the {\it plane},
with $\langle i|\ldots|j\rangle$
denoting $\langle\phi^\ast_i(\infty,\infty)\ldots\phi_j(0,0)\rangle$.
[To obtain~\ain\ one has to transform the correlators from the cylinder to
the plane; if $\phi_i$ is not
the ``planar'' state $|i\rangle$
appearing in the above has to be interpreted suitably, taking into account the
nontrivial transformation properties of $\phi_i$.]

As an easy consequence of~\ain\ note that all energy levels in the even
sectors of MTM and SGM must be identical. In particular, the SG and MT models
have  exactly the same finite-volume ground state energy $E_0(L)$.
(In these type of arguments we are making the very plausible assumption,
cf.~above, that equality in CPT means exact equality.)

Due to UV divergences the $a_{i,n}$ themselves are actually not well-defined
 when   $d=2-y\geq 1$, but since these divergences do not
depend on $\phi_i$ they cancel in the $\hat{a}_{i,n}$. See~\rourv\ for
more details about the UV divergences and their nontrivial effect on the
$e_i(\r)=(L/2\pi)E_i(L)$, as well as the regularization of possible
 IR divergences
of the $a_{i,n}$ (by analytic continuation in $d_i$) that is implicit in~\ain.

\subsec{Free MTM versus SG($\sqrt{4\pi}$)}

We now discuss in more detail the free MTM and the
SGM at its so-called ``free Dirac point'' SG$(\sqrt{4\pi})$, \ie~the
$V_{0,1}^{(+)}$ perturbations of the fermionic, respectively, bosonic
gaussian CFT of radius $r=1$.
The free MTM just describes a free massive
fermion $f$ and its antiparticle $\bar{f}$.
SG$(\sqrt{4\pi})$  contains only the soliton~$s$ and
the antisoliton~$\bar{s}$.  Our aim is to
find the {\it exact} finite-volume partition function of~\SGf.

\medskip
\no 4.3.1~~The Partition Functions

It is of course easy to write down the \pf\ of the free
MTM, eq.~(4.10) below.
The only nontrivial ingredient --- still easily
obtained --- is the finite-volume ground state energy; all energy gaps
just follow from the
dispersion relation of a free massive particle.
In contrast,
our derivation of the
partition function $Z_{{\rm SG}}$ of
SG$(\sqrt{4\pi})$
is not rigorous, and to motivate it we will first briefly discuss a different
pair of closely related but distinct (1+1)-dimensional QFTs whose
partition functions are rigorously known.  The theories in question are
the free massive Majorana fermion and the Ising field theory~\rIFT\rSMJ\ in its
high-temperature phase (IFT),
defined equivalently as a scaling limit of the lattice
Ising model (without a magnetic field) from above the critical temperature,
or as the   leading
thermal perturbation (with the appropriate sign of the coupling)
of the Ising CFT. We have discussed these theories in detail
before (see sect.~6 of~\rouriii\ and sect.~3.1 of~\rourviii) and therefore
can be brief here.

The  finite-volume partition function            of the IFT,
with periodic boundary conditions on the spin field,
can be
  derived by
taking a scaling limit of the Onsager solution of the Ising model on a finite
lattice (cf.~\rFerFi\rouriii).  It reads
\eqn\ZIFT{ \eqalign{Z_{{\rm IFT}}(\r) ~=~ {1\over 2}~
                                                        q^{e_+(\r)} ~ \Biggl\{
   & \prod_{n\in\ZZ +{1\o 2}} (1+q^{\eps_n(\r)})~ +
         \prod_{n\in\ZZ +{1\o 2}} (1-q^{\eps_n(\r)})  \cr
    +~ q^{\hat{e}_-(\r)} & \biggl[~ \prod_{n\in\ZZ} (1+q^{\eps_n(\r)}) -
           \prod_{n\in\ZZ} (1-q^{\eps_n(\r)})~ \biggr] \Biggr\} ~~.\cr } }
Here $q=e^{-2\pi/TL}$, $L$ is the ``volume'' of space,
$T$ the temperature,  $\r =Lm$ where
$m$
    is
the (infinite-volume) mass of the particle,
\eqn\epsn{ \eps_n(\r) ~=~ \sqrt{\Bigl({\r\o 2\pi}\Bigr)^2 + n^2} ~~,}
$\hat{e}_-(\r) =  e_-(\r) - e_+(\r)$, and~\rouriii\rourv\
\eqn\egsIFT{    e_\pm(\r)~ =~  -{\r\over 4\pi^2}\int_{-\infty}^\infty
    d\theta ~\cosh\theta~ \ln(1\pm e^{-\r\cosh\theta}) ~~.}
We have written everything in terms of dimensionless quantities.
The dimensionful
finite-volume energies $E_i(L)$ are read off from \ZIFT\ by expanding it
as $\sum_i e^{-E_i(L)/T}$, for instance
the ground state and first excitation energies are
$E_0(L)=(2\pi/L)e_+(\r)$ and $E_1(L)=(2\pi/L)e_-(\r) + m$, respectively;
the $(2\pi/L)\eps_n(\r)$ are just energy gaps of a free particle
of mass $m$ and momentum $2\pi n/L$.

Note that the first two terms in~\ZIFT\ (together
with the factor ${1\o 2}$) amount to a projection on states with an even
number of particles, including the vacuum,
the last two on odd-particle
states.
These
{\it even} and {\it odd sectors} are
distinguished by their
parity under
the $\ZZ_2$ symmetry of the theory, corresponding to the spin-reversal
symmetry of the Ising model.

For comparison recall that the partition function of the free massive Majorana
fermion (with anti-periodic boundary conditions) is
\eqn\ZMaj{Z_{{\rm Maj}}(\r)  ~=~
        q^{e_+(\r)} ~  \prod_{n\in\ZZ +1/2} (1+q^{\eps_n(\r)}) ~.}
The spectrum in the even sector, here with respect to the total fermion
number, is identical to that of the IFT, but the odd sectors of the two
theories are different.

The main point of~\ZIFT\ and \ZMaj\ is that their qualitative features are
exactly the same as those of the partition functions of the corresponding
UV CFT, obtained from the above by $m\to 0$.\foot{This gives the case of a
torus with perpendicular cycles, $q=\bar{q}$. Generalization of~\ZIFT, \ZMaj\
and other partition functions given below to an arbitrary torus amounts
to the replacement
$q^{\eps_n(\r)} \to
|q|^{\eps_n(\r)}~(q/\bar{q})^{{1\o 2} n}$,
explicitly identifying $n$ as the momentum quantum number.}
The building blocks of $Z_{{\rm IFT}}$ are the partition functions of a
free massive
Majorana fermion with the four possible boundary conditions in the time
and space directions,\foot{If
we denote by $(\alp,\beta)$, $\alp,\beta\in\{P,A\}$, the boundary conditions
(periodic or anti-periodic) in the time, $\alp$, and space, $\beta$,
 directions,
then the four terms in~\ZIFT\ correspond to $(A,A), (P,A), (A,P), (P,P)$,
respectively.}
and $Z_{{\rm IFT}}$ is obtained from $Z_{{\rm Maj}}$
by a GSO projection (cf.~sect.~2.2).
The only differences from the massless
case are the ground state scaling functions,
\ie~$e_+(r)$ in both sectors of the Majorana theory and
$e_\pm(\r)$ in the even/odd sector of IFT,
and the replacement of $|n|$ by the rescaled free particle
energies $\eps_n(\r)$.

\medskip
Because of a widespread confusion
in the literature,
we emphasize again that the IFT and the theory of a
free Majorana fermion are {\it not} identical.
The IFT describes an interacting boson created by the spin field $\sigma$,
with constant \sm\ $S=-1$~\rSMJ,
whereas the free massive Majorana theory describes,
of course,
a fermion with \sm\ $S=+1$. It
``just so happens'' that
the         partition function      of the IFT
with any allowed (see below) boundary conditions
on
$\sigma$
is identical to
some linear
 combination of ``partition functions'' of a free
Majorana fermion with various boundary conditions. We say ``partition
functions''      in quotes
because for a true partition function, namely the generating
function of the energy levels ${\rm Tr}\, e^{-\beta H}$, there is no choice for
the boundary conditions in the time direction. In a path integral formalism,
for example, one {\it must} choose anti-periodic boundary conditions
in the time direction
for fermions  and periodic ones for bosons,
otherwise one will get a different trace, \eg~${\rm Tr}\, (-1)^F e^{-\beta H}$
for a fermion with {\it periodic}     temporal        boundary conditions.
 The ``natural'' boundary conditions in the spatial direction
are the ones that are identical  to those in the time direction  --- then the
partition function will be invariant under exchange of space and time, and
presumably (cf.~sect.~5)
reveal the true     mutually local
 operator content of the theory --- although
one can also
impose various other        spatial        boundary  conditions
(depending on the symmetry of the theory).

\medskip
Some other features of \ZIFT\ will be illuminated as we now proceed to the
case of actual interest.
The \pf\ $Z_{\rf}(q,r$=$1)$
of the free {\it massless} Thirring model,
alias the free massless Dirac fermion,
can be written as $q^{-{1\o 12}}
  \prod_{n\in\ZZ +{1\o 2}} (1+q^{|n|})^2$, assuming $q\! =\!\bar{q}$ for
simplicity again.
We can also rewrite the \pf\ of the UV limit of SG($\sqrt{4\pi}$),
\eqn\ZSGmless{ Z_{\rb}(q,r=1) ~=~ q^{-{1\o 12}}~ \Biggl\{ ~
   \Bigl[ \prod_{n\in\ZZ +{1\o 2}} (1+q^{|n|})^2 \Bigr]_+ ~+~
   q^{{1\o 4}} \Bigl[ \prod_{n\in\ZZ} (1+q^{|n|})^2 \Bigr]_- ~\Biggr\} ~~, }
\ie~as a GSO projection of the massless Dirac fermion \pf. [Here and below
$[~\prod \ldots]_{\pm}$ denotes keeping only terms in the product with an
even/odd number of factors $q^{|n|}$ or $q^{\eps_n(\r)}$.]
The IFT-Majorana analogy suggests that the GSO mechanism extends to the massive
regime also in our case, namely that
the \pf\ of
SG$(\sqrt{4\pi})$ is related by a
GSO projection to that of the free massive Dirac fermion.
In this way
$Z_{{\rm SG}}$ will be automatically
modular invariant,
in particular invariant under the exchange $L \leftrightarrow 1/T$
(cf.~\rSalItz\ for the modular invariance of $Z_{{\rm IFT}}$).

The building blocks of \ZrmSG\ will therefore be ``partition functions''
 of a
free massive Dirac fermion with the four possible    boundary conditions.
These \pfs\ can be calculated either using $\zeta$-function regularization,
cf.~\rSalItz\ for the Majorana case, or, in a way that is simpler and allows
one to obtain more explicit expressions, by noting that
their
only nontrivial aspect
is the Casimir energy which depends on the
boundary conditions in the spatial direction. For a free theory the
finite-volume Casimir energy can be easily calculated using the
thermodynamic Bethe Ansatz (TBA), see \eg~\rouriii\ for details. For the
``natural'' boundary conditions of a free fermion, anti-periodic, each
fermionic degree of freedom (1 for the Majorana, 2 for the Dirac case)
contributes ${2\pi \o L} e_+(\r)$;
for periodic boundary conditions it is ${2\pi \o L} e_-(\r)$,  cf.~\egsIFT.

Putting everything together, the \pf\ of the free MTM, \ie~the free Dirac
fermion, is
\eqn\ZDirac{Z_{{\rm Dirac}}(\r)  ~=~
    q^{2 e_+(\r)} ~ \prod_{n\in\ZZ +{1\o 2}} ~ (1+q^{\eps_n(\r)})^2 ~~,}
and that of SG($\s4p$)
reads
\eqn\ZSG{ Z_{{\rm SG}}(\r) ~=~
     q^{2 e_+(\r)} ~ \Biggl\{ ~
   \Bigl[ \prod_{n\in\ZZ +1/2} (1+q^{\eps_n(\r)})^2\Bigr]_+ ~+~
  q^{2 \hat{e}_-(\r)} \Bigl[ \prod_{n\in\ZZ} (1+q^{\eps_n(\r)})^2\Bigr]_- ~
                                                           \Biggr\}   ~~. }

The only assumption used to obtain the above is that \ZrmSG\ is related by
a GSO projection to $Z_{{\rm Dirac}}$.
One nontrivial point in implementing this projection in the massive
case is the $[~~]_-$ projector --- as opposed to $[~~]_+$ ---
that we
used for
the second product in \ZSG. In the massless case the
two choices give the same result,
 because the \pf\ of a free {\it massless}
fermion with periodic boundary conditions in both
directions
vanishes identically. But in the massive case the two choices lead to
different physics; in particular,
if we use the  $[~~]_+$ projection the ground state is doubly degenerate and
no odd-particle states exist with periodic boundary conditions. This choice
is therefore not appropriate for the standard SGM; instead, we claim
that it gives the partition function of SG$(\s4p,2)$,
mentioned in sect.~3.\foot{The corresponding replacement of the minus sign in
front of the last term in~\ZIFT\ by a plus sign gives the partition function
of the theory obtained by taking a scaling limit of the Ising model from
{\it below} the critical temperature, see sect.~3.1 of~\rourviii\ for
details.}
The generalization to SG$(\s4p,k)$ for any $k$ will be discussed in~\rournext.

Though \ZSG\ is highly
plausible, we should provide
independent evidence for
its correctness.
We will do so below using
CPT, which indicates that
the small $\r$ behaviour of \ZrmSG\ is correct. First, however,
we want to discuss the ``IR interpretation'' of \ZrmSG,
which will also help us to see how conformal states in the UV CFT
``evolve'' into multiparticle states in the massive theory.
Note that
$Z_{{\rm Dirac}}$ is just the square of $Z_{{\rm Maj}}$, in accord with the
fact that the free massive Dirac theory describes the scattering of two,
not one, fermions with trivial diagonal \sm\ $S={\sl 1}$.
 Similarly, comparing
\ZrmSG\ and $Z_{{\rm IFT}}$ one expects that SG($\s4p$) describes the
scattering
of two {\it bosons} with diagonal \sm\ $S=-{\sl 1}$. We will
show that the
large $\r$ behaviour of \ZrmSG\ is
in perfect agreement
with this.

\bigskip
\no 4.3.2~~IR Check of \ZrmSG\

The first check of \ZrmSG\ is provided by the ground state energy, whose
rescaled form is $2e_+(\r)$. From the way \ZrmSG\ was obtained we know that it
is the Casimir energy of two free fermions of mass $m$ with anti-periodic
boundary conditions.
Fortunately --- and this is trivial to see with the TBA, cf.~\rouriii\ ---
a theory of two bosons of mass $m$ with diagonal \sm\ $S=-{\sl 1}$
has exactly the same ground state energy
with periodic boundary conditions,
in agreement with the above  picture of SG($\s4p$).
(Of course, we know independently from CPT that corresponding MT and SG
models have the same finite ground state energy for any $\beta$.)

It is well known~\rLui\rLuii\
that the \sm\ of any  (massive)
QFT determines the large-volume behaviour of its energy gaps.
The corrections
to the infinite-volume gaps (= sums of masses) have contributions that are
powerlike in $1/L$,
and others that fall off exponentially with $L$ (= the smallest extension of
the spatial volume).
These latter corrections
are due to off-shell effects,
basically
``virtual particles traveling
around the world''
(and     also  tunneling if the
QFT has a degenerate vacuum in infinite volume). For 1-particle states at
zero momentum
there are only exponential corrections and the leading ones can
be calculated in terms of the \sm\ \rLui\rouriv.

The powerlike dependence of
energy gaps of multi-particle states in any integrable
(1+1)-dimensional QFT can be determined exactly given its factorizable
\sm~\rLuii\rYZ\rLasMar\rourviii.
In the case of a diagonal and furthermore
constant \sm\ this is in fact basically trivial:
For an $N$-particle state in a theory with
$S=-{\sl 1}$ in finite volume $L$ with periodic boundary conditions,
the allowed
``single-particle momenta'' $p_j$ are determined by the quantization conditions
\eqn\BASGf{ e^{i p_j L} ~ (-1)^{N-1} ~=~ 1 ~, ~~~~~~~~~~~~j=1,\ldots,N. }
The momenta are therefore of the form $p_j = {2\pi \o L} n_j$, where
all $n_j \in \ZZ$ for an odd- and all
$n_j \in \ZZ +{1\o 2}$ for an even-particle state,
and in both cases the number of
 $n_j$ with the same value
must not exceed
the number of particle species
in the theory (because of the exclusion principle, cf.~subsect.~4.1).
Up to
exponentially small contributions,
the energy gap of a state in terms
of the ``momentum quantum numbers'' $n_j$ is
\eqn\ESG{\hat{E}(L) ~=~ {2\pi \o L} ~\sum_{j=1}^N ~\eps_{n_j}(m_j L) ~~.}
A look at \ZSG\ now reveals that up to exponential corrections its energy
levels,
as well as
their degeneracies, are exactly those of a theory of
two bosons of mass $m$ with
$S=-{\sl 1}$.

Note that in the even sector of    \SGf\
(and IFT)
there  are
{\it no} exponential corrections to \ESG, and in the odd sector the
corrections
are the same for {\it every} state,  namely,
for \SGf,
\eqn\FSMCexp{     {\Delta E(L) \o m} ~=~
       {2\pi \o m L}~ 2\hat{e}_-(m L) ~=~
       {4\o \pi}~ K_1(m L) ~+~{\cal O}(e^{-3 m L}) ~.}
This ``universality'' of the off-shell corrections is certainly not true
in a generic interacting QFT, and is a sign of how ``closely'' SG($\s4p$)
 and IFT
are related to free theories. More precisely, the vanishing of the
off-shell effects in the even sector of SG($\s4p$) and IFT is clear from the
fact that this sector is created by the        same fields
 that create the even sector in the free MTM and
the free Majorana theory, respectively.

It is not known how to calculate even the leading off-shell corrections for
an arbitrary multi-particle state in terms of the \sm,
except for 1-particle
states at zero momentum, \ie\ finite-volume masses.
For a (1+1)-dimensional
QFT with non-degenerate vacuum in infinite volume, containing only particles
of the same mass
$m$
and no poles in its scattering amplitudes, the finite-size
mass shift of a particle $a$ is (see eq.~(75) of~\rouriv)
\eqn\FSMCth{   {\Delta m_a(L) \o m}
  ~=~ -\int_{-\infty}^{\infty}
 ~{d\th\o 2\pi}~ e^{-m L \cosh\th} ~\cosh\th ~\sum_{b}
 \Bigl(S_{ab}^{ba}
                  (\th+{i\pi\o 2}) -1 \Bigr)  ~~+~{\cal O}(e^{-\sigma L})~~,}
with
$\sigma \geq \sqrt{3} m$. Therefore, a particle that scatters with itself
and exactly one other
mass-degenerate particle with scattering amplitudes
$S_{ab}^{ba}(\th)\equiv -1$
will have a finite-size mass shift of
{}~${\Delta m(L)\o m}
= 4 \int {d\th\o 2\pi} e^{-m L \cosh\th} \cosh\th
+{\cal O}(e^{-\sigma L}) = {4\o \pi} K_1(mL)+{\cal O}(e^{-\sigma L})$,
in   agreement with \FSMCexp.

\bigskip

\no 4.3.3~~CPT Check of \ZrmSG\

We now want to apply the CPT results~\eih\ain\ to check that the small $\r$
behaviour of \ZrmSG\ is consistent with the
formulation of SG($\s4p$) as
the $V_{0,1}^{(+)}$-perturbed bosonic gaussian model at $r=1$. To keep things
simple we will only consider levels in the zero momentum sector of the
spectrum that in the UV limit are created by
(spinless)
vertex operators that do not mix
under the perturbation
with other operators  of the same scaling dimension.
Since spin and  the electric charge $m$
are conserved by a $V_{0,1}^{(+)}$ perturbation, the
UV operators
$V_{m,0}$, $m\in \ZZ$,
obviously satisfy these criteria.
The only operators that $V_{0,\pm n}$, $n$$>$$0$, could couple to by a
$V_{0,1}^{(+)}$ perturbation are spinless descendants of
$V_{0,\pm (n-1)}$, but since $d_{0,n}-d_{0,n-1}=2n-1$
is odd, such operators cannot have the
same scaling dimension as  $V_{0,\pm n}$.  The $V_{0,\pm n}$ are therefore
part of a basis in the space of
$s$=$m$=$0$, $d$=$n^2$ fields
in which the
perturbation is diagonal, so that results of standard (non-degenerate)
perturbation theory apply.

\medskip
The first order term\foot{In the
following we denote $\hat{a}_{i,n}$ of subsect.~4.1 by $\hat{a}_n(\phi_i)$.}
 $\ah_1(A)$   vanishes for all
operators $A$ in the set $\{V_{\pm m,0}, ~V_{0,\pm n}\}$, $m,n\in \NN$.
 This means that the levels corresponding to the operators
$V_{\pm (2m+1),0}$, which are the only ones
in this set
 in the odd sector,
 must have exactly one term $\eps_0(\r)={\r\o 2\pi}$
in their energy, to cancel the ${\cal O}(\r)$ term in $2\hat{e}_-(\r)={1\o 4}-
{\r\o 2\pi} +{\ln 2\o 2 \pi^2} \r^2 + {\cal O}(\r^4)$. We will see that
this is indeed true after we have identified the ``IR labels'', namely the
momentum quantum numbers $n_i$ in \ESG,
of the levels corresponding to the above
operators. To identify these levels we proceed to $\hat{a}_2(A)$.

The correlators involved can be read off from
\Vcorr,
and the one complex
integral that has to be performed can be done either directly in polar
coordinates or using the ``generalized beta function'' of the complex number
field (cf.~\rbeta)
\eqn\genbeta{\eqalign{
  \int d^2z~ & z^{s+m/2} ~\bar{z}^{s-m/2}~(1-z)^{t+n/2}~(1-\bar{z})^{t-n/2} \
\cr
   &=   \veps
  \pi ~{\Gamma(s+{\scriptstyle {|m|\o 2}}+1)~
                          \Gamma(t+{\scriptstyle  {|n|\o 2}}+1)~
                          \Gamma(-s-t+{\scriptstyle {|m+n|\o 2}}-1)\o
                       \Gamma(-s+{\scriptstyle {|m|\o 2}})~
                          \Gamma(-t+{\scriptstyle {|n|\o 2}})~
                          \Gamma(s+t+{\scriptstyle {|m+n|\o 2}}+2)}~~.\cr}}
where $s,t\in {\bf C}$
(by analytic continuation from the region where the integral converges),
$m,n\in \ZZ$, and
$\veps=1$ if $mn\geq 0$ and $(-1)^{{\rm min}(|m|,|n|)}$ otherwise.

The results are
\eqn\aiiVo{\ah_2(V_{0,\pm n}) ~=~ {\kappa^2 \o 2} ~\biggl [
 \psi(n+{1\o 2})-\psi({1\o 2})\biggr] ~=~ \kappa^2
      ~\biggl (1 +{1\o 3}+{1\o 5} +\ldots  + {1\o 2n-1}\biggr) ~~,}
and
\eqn\aiiVm{\ah_2(V_{\pm m,0}) = {\kappa^2 \o 2} \biggl[
  \psi({m\o 2}+{1\o 2}) - \psi({1\o 2})\biggr] = \kappa^2
\cases{ \ln 2 + {1\o 2} + {1\o 4} +\ldots + {1\o m-1} ~~~m ~{\rm odd}\cr
  1 +{1\o 3}+{1\o 5} +\ldots  + {1\o m-1} ~~~~~~ m ~{\rm even}~, \cr}}
where $\psi(z) = {d\o dz} \ln\Gamma(z)$.

\medskip
To identify the corresponding levels as (multi-)particle states let us
label them
$s(n_1,\ldots,n_k)~\sb(n_{k+1},\ldots,n_\sN)$
 in terms of their momentum quantum numbers.
Because of the exclusion principle we can
restrict the quantum numbers to
$n_1 > n_2 > \ldots > n_k$ and $n_{k+1} > n_{k+2} > \ldots > n_\sN$, say.
Note that according to \ZrmSG\
the above
 state has UV
spin ~$s=\sum_{i=1}^\sN n_i$, and UV scaling dimension
{}~$d={1-(-1)^\sN \o 8} + \sum_{i=1}^\sN |n_i|$.

\medskip
$V_{\pm 1,0}$ are the lowest dimension operators above the vacuum and
one might expect that they correspond to the lowest
excited states also for finite $\r$,
namely
$s(0)$ and
$\sb(0)$, with scaled energy $2\hat{e}_-(\r)+\eps_0(\r)={1\o 4} +
{\ln 2\o 2\pi^2} \r^2 + {\cal O}(\r^4)$.
This          identification         is consistent with
\aiiVm, which furthermore allows us to conclude         that
\eqn\kapval{\kappa ~=~  {1\o \sqrt{2} \pi} ~~.}

\medskip
The identification of $V_{\pm 1,0}$ as soliton/antisoliton creation operators
in the UV is in agreement with our remarks in sect.~3,
showing that the electric
 charge that is still super-selected in a $V_{0,k}^{(+)}$
perturbation  should be identified as soliton number.
Using ~$\eps_n(\r)=|n|+{\r^2 \o 8\pi^2} {1\o |n|} +
{\cal O}(\r^4)$,
for
$n\neq 0$,
this identification             then              allows us to conclude
\eqn\Vmcorr{V_{\pm m,0} ~\leftrightarrow ~~\cases{
 ~s^\pm ({m-1\o 2},{m-3\o 2},\ldots,1,0,-1,\ldots,-{m-1\o 2})
        ~~~                              ~~   ~~~~ m~{\rm odd} \cr
 ~s^\pm ({m-1\o 2},{m-3\o 2},\ldots,{1\o 2},-{1\o 2},\ldots,-{m-1\o 2})
 ~~                                        ~~~~~~~~~ m~{\rm even}~~, \cr}}
which are the only states in the
charge $\pm m$
sector with the right UV
scaling dimension
{}~$d_{\pm m,0}$=$({m \o 2})^2$=${1\o 4} +
2 (1+2+\ldots+{m-1\o 2})$ for $m$ odd, and
{}~$d_{\pm m,0}$=$({m \o 2})^2$=$2 ({1\o 2}+{3\o 2}+\ldots+{m-1\o 2})$ for $m$
even.
[In the above $s^\pm$ stands for $s,\sb$, respectively.]

\medskip
The degenerate pair of levels corresponding to $V_{0,\pm n}$ is naturally
identified as the pair ~$s(n-{1\o 2},n-{3\o 2},\ldots,{1\o 2})
{}~\sb(-{1\o 2},-{3\o 2}, \ldots, -(n-{1\o 2}))$~ and ~$(s\leftrightarrow
\sb)$.
 From previous experience~\rLasMar\rourviii\ it is also plausible to conjecture
that, say, the pair
 ~$s(l+{1\o 2})\sb(-(l+{1\o 2}))$~ and ~$s(-(l+{1\o 2}))\sb(l+{1\o 2})$~
corresponds to certain
spinless
descendants of $V_{0,\pm 1}$ at (left and right) level $l$.
Finally, turning to sectors with nonzero momentum,
we conjecture that the special Virasoro
primaries in the vacuum sector of $d$=$\pm s$=$n^2$,
$n\in \NN$
 (which can be expressed in
term of the derivatives of $\phi$ and $\phib$, respectively, using Schur
polynomials~\rSchur)
correspond to the states
$s(\pm {1\o 2}, \pm {3\o 2},\ldots, \pm (n-{1\o 2}))
{}~\sb(\pm {1\o 2}, \pm {3\o 2},\ldots, \pm (n-{1\o 2}))$.

\bigskip
\subsec{Away from $\beta=\sqrt{4\pi}$}

In the previous subsections we
studied the SGM at $\beta=\sqrt{4\pi}$ in some detail.
We will
now briefly discuss what happens as $\beta$ moves below $\sqrt{4\pi}$, where
bound states   of the solitons,
so-called {\it breathers}, appear in the spectrum.
In particular, we will identify the field that
creates the first breather by looking at the finite-volume spectrum.

Consider the zero momentum sector of the spectrum of \SGf.
Recall that the soliton/antisoliton, whose rescaled energy gap
at rest is $\eps_0(\r)$, are created
by $V_{\pm 1,0}$ in the UV, and the lowest 2-particle states
$s({1\o 2})\sb(-{1\o 2}) \pm s(-{1\o 2})\sb({1\o 2})$ of energy
$2\eps_{{1\o 2}}(\r)$ by $V_{0,1}^{(\pm)}$.  Now lower $\beta$ by an
infinitesimal amount $\del$. The ``picture'' of the finite-volume spectrum,
\ie~the energy levels, can only change by an infinitesimal amount, but we
must now accommodate the first breather, a weak bound state of mass
$m_1=2m-{\cal O}(\del)$, in this picture. It is rather clear what will happen:
The bound state should correspond to the lowest of the former 2-particle
levels, the question only is which of $V_{0,1}^{(\pm)}$ creates it in the UV.
Using \genbeta\ one sees that $\hat{a}_2(V_{0,1}^{(\pm)})$
 is positive/negative,
so that presumably $V_{0,1}^{(-)}$ is the lowest level for all $\r$. We
therefore identify $V_{0,1}^{(-)}\propto \sin\beta\varphi$ as creating the
first breather for $\beta^2<4\pi$.

Initially this is justified only for small volume and $\beta$ just below
$\sqrt{4\pi}$. But besides the standard smoothness arguments, there are various
other reasons why this is very plausible. We note, for example, that
the scaling
dimension of $V_{0,1}^{(-)}$ drops below that of $V_{\pm 1,0}$ exactly when
the mass of the first breather drops below that of the solitons.\foot{This
happens at $r=1/\sqrt{2}$, the self-dual point, where the bosonic gaussian
model has a level one $\widehat{su}(2)$$\times$$\widehat{su}(2)$
Kac-Moody symmetry (see~\eg~\rWZWFZ).
  Under a $V_{0,1}^{(+)}$ perturbation a global $SU(2)$
survives,  with respect to  which the
conformal fields $V_{\pm 1,0}$ and $V_{0,1}^{(-)}$ --- and thus, according to
our identifications, also the
soliton/antisoliton and the first breather --- form a
triplet. This implies that their finite-volume levels must be exactly
degenerate for {\it all} $L$, and that the SG amplitudes at $\beta^2=2\pi$
should be invariant under permutations of
the three particles. This is indeed true for our sign of the SG \sm. ~It was
argued on different grounds in~\rColeSchw\ (cf.~also~\rMWsusy) that
SG$(\sqrt{2\pi})$ has an $SU(2)$ ``isospin'' symmetry. The argument involves
viewing SG$(\sqrt{2\pi})$ as a subtheory of the $SU(2)$ Schwinger model in a
certain limit. The solitons turn out to be ``quark-antiquark'' bound states
in terms of the Schwinger model fermions. Therefore they should be bosons!
(This was not commented upon in~\rColeSchw.)}
More convincingly,
  standard
perturbation theory around
  the free massive theory
$\beta=0$ provides
quantitative evidence~\rAraKor\rDaHaNe\ that $\varphi$ creates the first
breather, which is in any case the only plausible candidate for what it could
create. Of course, from a non-perturbative point of view $\varphi$ is
compactified, \ie~not really well-defined. The simplest well-defined field
that creates the same asymptotic state as $\varphi$, that is, gives rise to
the same  \sm, is $\sin\beta\varphi$. (See chapter~2 of~\rColeBook\ for a
discussion of when two different fields lead to the same \sm.)

Recall from sect.~4.1 that the first breather, and as we now know
$\sin\beta\varphi$, are $C$ and $P$ odd.
By just looking at the SG lagrangian
it seems that one can simply
{\it choose}  $\sin\beta\varphi$ to have any $C$ and $P$ parity. On a
non-perturbative level this is not the case, though, because
$\sin\beta\varphi$
creates a particle that is a bound state of the underlying solitons, and its
parity properties must be consistent with the dynamics of these solitons.
Similarly, one can argue that for the other $V_{0,k}^{(+)}$  ($V_{k,0}^{(+)}$)
perturbations the SG ``field'' $\varphi \propto \Phit ~(\Phi)$ is a
pseudo-scalar
(cf.~\rournext). This is not just true for the perturbed bosonic gaussian CFTs,
but also for the fermionic ones, where it is in fact obvious: There we have
direct access to the ``underlying'' fermions and by bosonization the fermion
current $J^\mu \equiv \, : \! \Psib \gam^\mu \Psi \! :\,
 \propto \eps^{\mu\nu} \partial_\nu
\varphi$. The fact that $J^\mu$ is a current, not a pseudo-current, implies
that $\varphi$ must be a pseudo-scalar.

We conjecture that the $n$-th breather is created by $V_{0,n}^{{\scsc((-)^n)}}$
in the UV limit. Since the
$n$-th breather
can be interpreted as
a bound state of $n$ lightest breathers~\rDaHaNe,
this is in accord with the fact that a suitably defined ~$:(V_{0,1}^{(-)})^n:$~
equals ~$V_{0,n}^{{\scsc ((-)^n)}}$.

Finally we should comment on what looks perhaps a bit strange in the above
sketched picture of the finite-volume spectrum as $\beta^2$ goes below $4\pi$.
Namely, the energy level which at $\beta^2 \! =\! 4\pi$ corresponds to
$s({1\o 2})\sb(-{1\o 2}) - s(-{1\o 2})\sb({1\o 2})$,
``belongs'' to the breather
when $\beta$ is just infinitesimally smaller.
But there is still a lowest
anti-symmetric 2-particle state; what is its energy? The only possibility
is that its
energy gets ``bumped up'' to $2\, \eps_{3/2}(\r)$. [Similarly, one
of each of the two zero momentum 2-particle levels  of energy
$2\, \eps_{n+{1\o 2}}$ must be bumped up to $2\, \eps_{n+{3\o 2}}$,
$n=1,2,\ldots$.]
Naively it looks as if the spectrum changes discontinuously.
However, an infinitely
weak bound state in finite volume is not distinguishable from an unbound
2-particle state, so that there is no observable discontinuity at
$\beta^2\! =\! 4\pi$; sets of energy levels of given symmetry properties change
completely smoothly. Note also that there is nothing wrong with the fact
that the (zero momentum) breather starts its existence with
energy ${2\pi\o L}\,           2\,
\eps_{{1\o 2}}=\sqrt{(2m)^2 + ({2\pi\o L})^2}$, which
differs at ${\cal O}(L^{-2})$ from $2m$, whereas finite-size mass
corrections are supposedly    exponentially small in $L$. The point is that
for a weak bound state of mass  $m_1=2m-\del$, $\del \ll 1$, all we know is
that the  finite-size mass corrections are ${\cal O}(e^{-\del m L})$
for $\del m L \gg 1$~\rLui\rouriv;  when $\del$ is infinitesimal they can
certainly be  ${\cal O}(L^{-2})$ for all $L$.

\newsec{Discussion and Further Examples}

In retrospect  the inequivalence  of the SG and MT models is rather obvious.
Let us briefly summarize the main points and why, we think, this was not
noticed much earlier. The equality of correlation functions in certain
sectors of SGM and MTM appeared to be good evidence for SGM=MTM at a time
when not many examples of distinct QFTs with nontrivial subsectors of
identical operators were known. With the advent of ``modern CFT'' many such
examples were discovered, but then the emphasis on {\it full} modular
invariance (due to the influence of string theory) prevented an appreciation
of the significance of the {\it fermionic} gaussian models. Furthermore,
the fact that various QFTs can be constructed out of the same compactified
boson was often believed to mean that these theories are ``equivalent''; but,
if they contain operators that are not mutually local, they
must
be distinct.  It is also important to realize that in {\it the} (or at least
one) right way of looking at the SG and MT models the SG ``field''
$\varphi$ is compactified. This is crucial
for understanding
the ``higher harmonic''
analogs of the SG and MT models, where several periods of the cosine
potential fit on the circle on which $\varphi$ lives, so that these models
are ``kink theories'' with a degenerate vacuum in infinite volume.

Our story of course also has a statistical mechanics version. Among lattice
models whose critical points are described by $c$=1 CFTs one must not only
distinguish between models in the orbifold
and bosonic gaussian families of universality classes, but also the latter
from the fermionic gaussian one. This is more subtle, since the bosonic
and fermionic gaussians have the same generic $O(2)$$\times$$O(2)$ symmetry,
and even a quantitative measure like the
free energy per site in the thermodynamic limit
will not distinguish between these universality classes
(the same situation arises
in the case of IRF lattice models related by ``orbifolding''~\rFenGin).
One needs critical exponents corresponding to
fermionic operators or (subleading) eigenvalues of the transfer matrix
corresponding to levels in the ``odd sectors'' to distinguish them. Some
remarks on consequences of our results for various scaling limits of the
spin ${1\o 2}$ XYZ chain are given in appendix~B.

The bosonic and fermionic gaussian CFTs are related by
a   twist (followed by a projection), and
we discussed the various repercussions this has in the perturbed theories.
Our analysis should be extended to other
pairs of non-scale-invariant QFTs defined as ``the same'' perturbation of CFTs
related by twists,
alternatively a (generalized) orbifold construction.
There are numerous such examples, and we will now briefly
discuss several of them, leaving details for future work.

One class of examples was already encountered in sect.~3, namely the
$V_{0,k}^{(+)}$ perturbations, $k\in \NN$, of the gaussian CFTs. We note that
the operator $V_{0,k}^{(+)}=\sqrt{2}\cos(2kr\Phit)$
at $r$ is identical --- in terms of its behaviour in
correlation functions --- to $V_{0,1}^{(+)}$ at radius $r'=kr$. Moreover,
the gaussian CFT (bosonic or fermionic) at $r$ can be obtained
from the one at  $r'$ by a $\ZZ_k$-twist~\rDFMS\rGinspLH, the $\ZZ_k$
being the discrete subgroup of the $U(1)$ symmetry (see subsect.~2.1)
of the model at $r'$, generated by $(\Phi,\Phit)\to  (\Phi+2\pi r'/k,\Phit)$.
We may therefore think of the theories
$A_{\rb,\rf}(r,V_{0,k}^{(+)})$ with $k\geq 1$
as $\ZZ_k$-``massive orbifolds'' of
$A_{\rb,\rf}(kr,V_{0,1}^{(+)})$ (or {\it vice versa}; then
the $\ZZ_k$ is the subgroup of   the  $\Ut(1)$  that  is not broken by   a
$V_{0,k}^{(+)}$ perturbation). This point of view is rather useful
in practice. For instance, it allows us to obtain~\rournext\
the full exact finite-volume spectra and \sms\
of the kink theories
$A_{\rb,\rf}(1/k,V_{0,k}^{(+)})$
by twisting those of
$A_{\rb,\rf}(1,V_{0,1}^{(+)})$,
the latter being
the SGM at $\beta^2=4\pi$ and the free MTM,
respectively.

Another class of examples is provided by perturbations of the
$c$=1 orbifold CFTs (see \eg~\rGinspLH).
These CFTs are related to the bosonic gaussian ones through a $\ZZ_2$ twist,
in this case with respect to $R \tilde{R}\! :~(\Phi,\Phit)\to (-\Phi,-\Phit)$.
They are bosonic, \ie~do not contain any fields of half-odd-integer Lorentz
spin, and
just have
 a discrete ${\bf D}_4$ global symmetry (the symmetry
group of the square).
Contrary to $V_{m,0}^{(-)}$ and $V_{0,n}^{(-)}$, which are projected out,
the operators $V_{m,0}^{(+)}$ and $V_{0,n}^{(+)}$ are
still
part of the
orbifold operator algebra (OPA),
and can be used to generate integrable perturbations.
However, the sign of the perturbation now matters in contrast to the
gaussian case, the two     perturbations
being related by the
continuum analog of the Kramers-Wannier duality of the underlying
Ashkin-Teller
model.
The corresponding theories
$A_{{\rm orb}}^{(\pm)}(r,V_{0,k}^{(+)})$, $k=$$1,2,\ldots < \sqrt{2}/r$,
have to be studied separately, as they differ in their kink-structure
and hence also their \sm\ and finite-volume spectrum~\rourviii.

A particular case of the above perturbed orbifold CFTs that has
already   been    analyzed in some detail~\rourii\rouriii\ is that of
$A_{{\rm orb}}^{(+)}(\sqrt{{N\o 2}},V_{2,0}^{(+)})$, $N=2,3,\ldots$.
[The sign choice for the direction of the perturbation is a convention.
We take the plus sign to correspond to the case where the vacuum of the
theory is non-degenerate in infinite volume; in the opposite direction it
is doubly degenerate.]
These
theories have been argued~\rourii\ to
be described by the so-called $D_\sN^{(1)}$ diagonal \sm\ theories, which can
be thought of as massive orbifolds of the SGM at the points
$\beta^2=8\pi/N$ where soliton-antisoliton scattering is reflectionless.
We pointed out
that there are sign differences
 between the scattering amplitudes of the      two
``fundamental particles'' (which are bosons) in the $D_\sN^{(1)}$ theory
and those of the solitons in the corresponding SGM.\foot{However,
when writing~\rourii\ we
still had the wrong impression that the latter are identical to
the amplitudes of the fermions of the MTM.}
For instance, in the $N=2$ case
the theory
contains only the two
self-conjugate
fundamental particles, with the  nonzero amplitudes
$S_{11}^{11}=S_{22}^{22}=-1=-S_{12}^{21}=-S_{21}^{12}$.
Thus $A_{{\rm orb}}^{(+)}(r=1,V_{2,0}^{(+)})$
clearly describes two decoupled copies of the Ising field theory,
and the \pf\ is $Z_{D_2^{(1)}}(\r)=Z^2_{{\rm IFT}}(\r)$, cf.~\ZIFT.

It is natural to look for the full one-parameter family of factorizable \sms\
describing $A_{\rm orb}^{(+)}(r,V_{2,0}^{(+)})$ for arbitrary $r$, and
see if and how they are related to the \sm\ of SG($\beta$=$\sqrt{4\pi}/r$),
namely
$A_\rb({r\o 2},V_{1,0}^{(+)})$.
The  \sm\ should be {\bf D}$_4$-symmetric for any $r$, since this
symmetry of the UV CFT is not broken by the perturbation.
A two-parameter family of {\bf D}$_4$-symmetric ``elliptic $S$-matrices'',
formally describing the scattering of a doublet of particles,  was
in fact constructed by Zamolodchikov~\rZamDiv. It is now believed that
elliptic solutions to the bootstrap constraints are not
relevant for QFT,
\ie~only their degenerate trigonometric limits may describe
relativistic QFTs. Zamolodchikov noticed that in one trigonometric limit,
$l\to 0$ in his notation, the {\bf D}$_4$ symmetry is enlarged
to $O(2)$ and
the \sm\ reduces to that of the SGM (or MTM, depending on the overall sign).
But there is
another limit, $l\to 1$,
which is not $O(2)$-symmetric!
In this limit the  nonvanishing \smes\ for the
doublet of
self-conjugate
particles
are
\eqn\orbSM{\eqalign{
  S_{11}^{11}(\th)& ~=~ S_{22}^{22}(\th) ~=~ {1\o 2}~(+s+t+r)(\th) ~,\cr
  S_{11}^{22}(\th)& ~=~ S_{22}^{11}(\th) ~=~ {1\o 2}~(-s+t+r)(\th) ~,\cr
  S_{12}^{21}(\th)& ~=~ S_{21}^{12}(\th) ~=~ {1\o 2}~(-s-t+r)(\th) ~,\cr
  S_{12}^{12}(\th)& ~=~ S_{21}^{21}(\th) ~=~ {1\o 2}~(+s-t+r)(\th) ~, \cr}}
in terms of the SG
$ss$ and $s\bar{s}$ transmission and reflection
amplitudes $s$, $t$ and $r$, respectively (with the correct sign).
This is our conjecture for the exact \sm\ of the fundamental particles of
the theory  $A_{\rm orb}^{(+)}(r\! =\!\sqrt{4\pi}/\beta,V_{2,0}^{(+)})$.
It seems to be true in general, cf.~\rFenGin\rZamkink, that the \smes\ of
an orbifold theory are linear combinations of those in the ``original''
theory.
Note that when $\beta=\sqrt{8\pi/N}$, $N=2,3,\ldots$,
the amplitudes \orbSM\ coincide with those of the
$D_\sN^{(1)}$
scattering theory
(in a 1-particle basis where the fundamental particles are self-conjugate
even for $N$ odd, cf.~\rourii).
The bound state amplitudes can be obtained via the bootstrap;
the number and masses
of the bound states are exactly
as in the corresponding SGM, but their interpretation as composites of
the   fundamental particles is different.

The other ``thermal'' perturbations of the orbifold CFTs
deserve further investigation.  In addition, there are ``magnetic''
perturbations~\rHenSal\rHenLud\
which have no analog in the gaussian case, being induced by
twist
operators that cannot be expressed as exponentials of a free boson.

\medskip
Finally, we would like to discuss theories that can be obtained by ``fermionic
twists'' from bosonic models.
To gain some insight about
such theories,
consider first the CFTs at the
UV limits of the SGM and its fermionic partner, the MTM.
As discussed in sect.~2,
the corresponding \pfs\
$Z_{\rb,\rf}(q,r)$
seem to cover
all $\Gamma'$-invariant
partition functions of $c$=1 unitary CFTs having (at least) $U(1)$ symmetry.
[By this statement we mean the following. Let $\chi_\Delta(q)$ be
the character of the Virasoro irrep of central charge $c$=1 and highest weight
$\Delta \geq 0$.
We conjecture that the only
{}~$Z(q)=\sum_{(\Delta,\bar{\Delta})} {\cal N}_{\Delta,\bar{\Delta}}~
\chi_\Delta(q) \chi_{\bar{\Delta}}(\qb)$~
with ${\cal N}_{\Delta,\bar{\Delta}}\in \ZZ_{\geq 0}$,
${\cal N}_{1,0} \geq {\cal N}_{0,0} =1$ satisfying
$Z(q)=Z(e^{4\pi i}q)=Z(\tilde{q})$, are $Z_{\rb,\rf}(q, r)$ for some $r$.]
We are not aware of a proof
of this statement
(partial support, related to the particular case of full $\Gamma$-invariance,
can be found in~\rKir\rDVV). In any case,
our discussion (sect.~2)
of consistent
OPAs       for     $U(1)$$\times$$U(1)$-symmetric CFTs     suggests a
deep general connection between (sub)modular invariance of the partition
function and local properties of the corresponding OPA.
Namely, given a $\Gamma'$-invariant partition function, there appears to
exist a (not necessarily unique)
consistent OPA --- in particular satisfying mutual locality in a
``maximal'' way --- giving rise to that partition function.

Such a connection is alluded to in many places in the CFT literature
(see \eg~\rGepn\rTB)
though
always, it seems,
restricted to the case of full modular invariance.
We here draw attention to the more general case of
$\Gamma'$-invariance whose consequences
do not seem to have been systematically            explored so far.

\medskip
In some cases the construction of $\Gamma'$-invariant fermionic
CFTs is rather trivial. For example, the Neveu-Schwarz sector
(including the fermionic components of the super-fields) of any
super-CFT fits the bill. A bit more generally,
one can take the ``Neveu-Schwarz sectors'' of
CFTs invariant under (not necessarily supersymmetric) fermionic
$W$-algebras~\rNahm, namely chiral algebras containing
chiral
fields of half-odd-integer spin. However, this class
of theories does not
automatically
 cover CFTs that contain fermionic
fields which are not chiral. To discover such CFTs using ``chiral techniques''
one apparently has to consider twisted bosonic $W$-algebras (see
footnote~24 below).

Alternatively,
putting aside for
a
moment the question of full consistency of the CFT,
one might attempt to simply classify
$\Gamma'$-invariant
\pfs\ built of characters of
the Virasoro or any other extended chiral algebra in
the same way as full modular invariance  has been
analyzed.
For Virasoro minimal models
 this constraint is highly restrictive. Although we do not
have a complete proof at this point, we believe that there exist only
two series of ``fermionic'' solutions,
one of them exceptional (to be contrasted with
five series in the bosonic case~\rCIZ, three of which are exceptional).
Explicitly, in the notation of~\rCIZ,
the $\Gamma'$-
(but not $\Gamma$-)
invariant \pfs\ we find read
\eqn\fVir{ \eqalign{
  p'=4\r~~(\r \geq 1)~~~~~~~~~~~~~&
  {1\o 4}~\sum_{s=1}^{p-1}~\sum_{r~{\rm odd}=1}^{p'-1}
          |\chi_{rs}+\chi_{r,p-s}|^2    \cr
  p'=4\r +2~~(\r\geq 1)~~~~~~~~&
  {1\o 2}~\sum_{s=1}^{p-1} ~\Biggl\{ \sum_{r~{\rm odd}=1}^{p'-1}|\chi_{rs}|^2
   + \sum_{r~{\rm even}=2}^{2\r}(\chi_{rs}\chi_{r,p-s}^* + c.c.) \Biggr\}   \cr
  p'=12~~~~~~~~~~~~~~~~~~~~~~~~~&
  {1\o 2}~\sum_{s=1}^{p-1}
          |\chi_{1s}+\chi_{5s}+\chi_{7s}+\chi_{11s}|^2 ~~.   \cr} }
We will refer to the \pfs\ on the first two lines as members of the
{\it fermionic} $D$-{\it series}, the ones on the last line as the
{\it fermionic} $E_6$-{\it series}.
As in the bosonic case, the {\it unitary fermionic series}
of central charge $c=1-{6\o m(m+1)}$ corresponds to $p=m,p'=m+1$ or
$p=m+1,p'=m$, with $m=3,4,\ldots$. There is one model in the unitary
fermionic $D$-series for each $m=3,4,\ldots$, and
two more unitary  fermionic models are found in the $E_6$-series,
\ie~at $m=11,12$.

Let us make some brief comments on \fVir:

\no {\bf (i)} As discussed above, given a $\Gamma'$-invariant as in \fVir\
does not quite guarantee the existence of a consistent CFT having that
invariant as a \pf. In particular, one has to check that an OPA whose
``content'' only is encoded in the invariant satisfies all required
properties. First note that invariance under
the modular transformation $T^2$ implies $s\in {1\o 2}\ZZ$
for all fields in the model.
Next, closure of all the OPAs encoded in \fVir\
is easily verified using the (chiral) fusion rules
of the minimal models~\rBPZ, ammended with conservation of $(-1)^F$
($=1$ for fields of spin $s\in \ZZ$, $-1$ for $s\in \ZZ+{1\o 2}$).
Conservation of $(-1)^F$ also ensures mutual locality.
However, associativity is much more difficult to prove,
requiring essentially the calculation of all OPE coefficients.
This problem in the fermionic case, which      by definition
involves a
non-diagonal submodular invariant \pf, is similar to that of computing
OPEs in non-diagonal bosonic CFTs.  A satisfactory method to solve this
problem in both cases still remains to be found (cf.~\rourvii\ and references
therein). We are nevertheless rather confident that all the \pfs\ listed
in \fVir\ do correspond to consistent CFTs.

\no {\bf (ii)} The fields of $(-1)^F=1$ in the  model labeled by
$(p,p')$ in the fermionic $D$-series
coincide with the $\ZZ_2$-even fields in the corresponding bosonic
$A$- and $D$-model, and the same relation holds between corresponding
fermionic and bosonic $E_6$-models.\foot{Cf.~\rCIZ\rGepn\ for~the~$\ZZ_2$
symmetry~in~the~bosonic models,~present~in~the~$A$,~{$D$,} and $E_6$-models
when one of $p,p'$ is even. The fact that
$p'$ must be even in the first two lines of \fVir, allowing for
the $\ZZ_2$ symmetry that any fermionic model must obviously have, motivated us
to call the corresponding series of theories the $D$-series;
the bosonic $A$-series
contains also models with $p,p'$ both odd,
which have no fermionic counterparts.}
This suggests that the
series involved
are related by
$\ZZ_2$-twists.

\no {\bf (iii)} The fermionic $D$-series model $(p,p'=4\r)$ is minimal
and diagonal with respect to the fermionic $W(2,\Delta_{1,p-1})$ chiral
algebra, in the notation of~\rNahm, where $\Delta_{1,p-1}=(p-2)(p'-2)/4 \in
\NN-{1\o 2}$. Similarly the fermionic models in the $E_6$-series are
minimal and diagonal with respect to $W(2,{p\o 2}-2,p-3,{5(p-2)\o 2})$.
(In fact, some of the corresponding $\Gamma'$-invariants have been presented
in~\rNahm.)
The sums of Virasoro characters
$\chi_{rs}+\chi_{r,p-s}$
and $\chi_{1s}+\chi_{5s} + \chi_{7s}+\chi_{11s}$
which appear on the first      and third
line
of \fVir,    respectively,   are characters of the above
chiral algebras.

\no {\bf (iv)} The unitary fermionic model $m$=3 is just the free massless
Majorana fermion, and the $m$=4 model is the
first model in the $N$=1 superconformal unitary series
(containing only the Neveu-Schwarz sector),
from which the
tricritical Ising CFT is obtained by a fermionic twist.
These two models are special in that their global symmetry is
$\ZZ_2$$\times$$\ZZ_2$, whereas all other models in the fermionic
unitary series are only $\ZZ_2$-symmetric.
This is due to the fact that for $m$=$3,4$ all fields have
left and right
Kac labels of the form $(n,1)$   or  $(1,n)$, so that one can define
separate left and right charges  $(-1)^F$ and $(-1)^{\bar{F}}$,
not just their product $(-1)^F=(-1)^{2s}$.
    The
action
    of each one of them
on the
    {\it bosonic}
fields in the models has the same effect as
that of the Kramers-Wannier duality transformation in the corresponding
bosonic CFTs. But while duality maps the full local operator algebra of
the bosonic CFTs onto a {\it different} algebra, exchanging order and
disorder operators (cf.~\eg~appendix E of~\rBPZ), it is    promoted
to a global $\ZZ_2$ symmetry in the fermionic  $m$=$3,4$ models.
These observations become important when considering perturbations of
these CFTs, see below.

\no {\bf (v)} The unitary fermionic models with $m\geq 5$
are apparently new. We now discuss
in some detail
the $m$=5 case,
the fermionic
partner of the bosonic tetracritical Ising ($A$-series) and 3-state
Potts ($D$-series) CFTs.
Note that the \pf\ of the fermionic $m$=5 model can be rewritten as
\eqn\fpotts{ Z_{\rf}^{D_5}
 ~=~ Z_{\rb}^{A_5}-(|\chi_{1,2}-\chi_{4,2}|^2 + |\chi_{2,2}-\chi_{3,2}|^2)~~,}
where $Z_{\rb}^{A_5}$ is the \pf\ of the tetracritical Ising CFT,
in a self-explanatory notation. (Formulas similar to \fpotts\
exist
for all the \pfs\ on the second line of \fVir.)
The expression in parenthesis in \fpotts, evidently $\Gamma'$-invariant
by itself, is~\rMICar\ the ``\pf'' $Z_{\rb}^{D_5}(T,T)$ of the 3-state
Potts CFT with $\ZZ_2$-twisted boundary conditions on the complex
spin field $\sigma$ of the model
(namely, $\sigma(z+\omega_i,\zb+\omega^*_i)=\sigma(z,\zb)^*$
for $i=1,2$, where $\omega_i$ are the two periods of the torus).
In this notation, {\it the} \pf\ of the 3-state Potts CFT is
$Z_\rb^{D_5}=Z_\rb^{D_5}(P,P)$, where $P=~$periodic.
Moreover, we note     that
\eqn\mvproj{ {1\o 2}~\bigl[ Z_\rb^{D_5}(P,P) + Z_\rb^{D_5}(P,T) +
    Z_\rb^{D_5}(T,P) \pm Z_\rb^{D_5}(T,T) \bigr] ~=~
    \cases{ Z_\rb^{A_5}  \cr  Z_\rf^{D_5}  \cr}}
(cf.~\rMICar\ for the upper case),
showing
 how the tetracritical
Ising and fermionic $m$=5 CFTs are obtained by twisting the 3-state Potts
CFT.\foot{Recall that the 3-state Potts CFT, \ie~the bosonic $D_5$ model,
is minimal and diagonal with respect to the $W_3$ algebra.
It was noted in~\rKiZh\ that
the combinations of Virasoro characters $\chi_{1,2}-\chi_{4,2}$ and
$\chi_{2,2}-\chi_{3,2}$
appearing in \fpotts\ are essentially $T$-transformed
characters of the {\it twisted} $W_3$ algebra,
suggesting that this twisted chiral algebra plays a role in both the
bosonic $A_5$ and fermionic $D_5$ models.}
In terms of the
$m$=5 nonlocal ``theory''
 containing the union of fields in the bosonic $A_5$, $D_5$ and
fermionic models,
the $\ZZ_2$-odd $D_5/A_5$ fields can be thought
of~\rMICar\ as order/disorder operators, their (double-valued)
OPEs generating the fermionic $(-1)^F=-1$ fields. This is similar to
the $m$=3 case,
but in the case
at hand the fermions are not free, having conformal dimensions
$({1\o 40},{21\o 40})$, $({21\o 40},{1\o 40})$, $({1\o 8},{13\o 8})$,
and $({13\o 8},{1\o 8})$.

\medskip
Given
some new fermionic CFTs, the next step is to consider their relevant
perturbations.
Since we restrict ourselves to spinless perturbations, where the perturbing
field is in the $(-1)^F$=1 sector of the theory, such fermionic perturbed
CFTs necessarily have
bosonic partners where ``the same''
perturbing field is in the even sector of the unperturbed bosonic CFT.
The difference between the fermionic and bosonic theories
is expected to be in general more drastic  than
in the case of the MT {\it vs.}~SG models,
where even though
the operator algebras of the two theories are
different,
the particle spectrum
is identical and there
are only sign differences in certain \smes.

One particularly interesting
family of theories is the $\phi_{1,3}$-perturbed models in the unitary
series.
Perturbations in the massive direction of the bosonic $A$-series models,
known to lead
 to the {\it restricted sine-Gordon} (RSG) models~\rRSG,
describe the scattering of a multiplet of (bosonic) kinks~\rZamkink.
(The simplest such kink theory, the $m$=3 case,  is just the IFT in the
{\it low}-temperature phase~\rourviii.)
Correspondingly, one might expect the $\phi_{1,3}$-perturbed fermionic models
to describe theories of fermionic kinks.

For $m>3$ this is presumably true. For $m$=3, however,
the enlarged symmetry of the theory (see (iv)
above) shows that the sign of the perturbation is irrelevant --- the sign of
the fermion mass term in the lagrangian is just a convention ---
so
there is no kink phase.
For higher models in
the fermionic series the direction of the perturbation does matter,
presumably leading to a kink phase in one direction.
It
is rather clear that in this phase the degeneracy of the
vacuum and therefore the kink structure will differ from that of the
corresponding RSG models
(implying, in particular, that the \sms\ of the perturbed bosonic and
fermionic theories will differ by more than just signs).
In fact, since the $m-1$ degenerate vacua in the $m$-th RSG model
are presumably
 created by the spinless fields $\phi_{rr}$
($r=1,2,\ldots,m-1$)~\rourv\ in the bosonic UV CFT, there are only $[{m\o 2}]$
vacua in the perturbed fermionic CFT, created by the even fields
$\phi_{rr}$ ($r=1,3,\ldots,2[{m\o 2}]-1$) that survive the twist.
The relation between these massive theories
is therefore
similar to that between $SG(\beta,k)$ with $k\in 2\NN$
and its fermionic partner
(see sect.~3 and~\rournext),
 and requires
further investigation.\foot{A first attempt to
find the \sm\ for
the $m$=4 case was made in~\rFenFSM.}
In the {\it massless} direction
of the $\phi_{1,3}$ perturbation we expect RG flows~\rZamflow\rLudCar\
 between the fermionic unitary CFTs $m$ and $m-1$ ($m=4,5,\ldots$ in the
fermionic $D$-series, $m=12$ in the $E_6$-series), the reasoning
being along the same lines as in~\rourvii.

\medskip
\bigskip
{\vbox{\centerline{\bf Acknowledgements}}}
\medskip
We would like to thank T.~Banks, M.~Douglas, P.~Fendley, D.~Friedan,
V.~Korepin, A.~LeClair, B.~McCoy, M.~Ro\v cek, K.~Schoutens, R.~Shrock,
D.~Spector, H.~Tye, E.~Verlinde, and S.~Yankielowicz
for discussions. The work of T.R.K.~is supported
by NSERC and the NSF, and that of E.M.~by the NSF, grant no.~91-08054.

\bigskip

\appendix{A}{Conventions}

When working in Minkowski space $x^\mu$, $\mu=0,1$, our signature is $(+,-)$.
For the Dirac matrices, $\{\gam^\mu,\gam^\nu\}=2g^{\mu \nu}$, we use the
representation
\eqn\Dirmats{ \gam^0 ~=~ \pmatrix{0 & 1\cr 1 & ~0} ~~, ~~~~
              \gam^1 ~=~ \pmatrix{0 & 1\cr -1 & 0}~~,~~~~
              \gam^5 ~=~ \gam^0 \gam^1~=~\pmatrix{-1 & 0\cr 0 & 1}~~.  }
Wick rotation to euclidean space
corresponds to
$(x^0,x^1)\to
(x_1=x^1,x_2=ix^0)$, and complex euclidean coordinates are defined by
$(z,\zb)={1\o 2}(x_2+ix_1, x_2-ix_1)$.

For the free massless scalar field of the gaussian CFT we adopt the
normalization (and choice of regulating mass) used in~\rGinspLH,
except that we denote the field by $\Phi(z,\zb)$ instead of $X(z,\zb)$.
Explicitly, the 2-point function is
\eqn\twoptfc{ \la \Phi(w,\wb)\Phi(z,\zb)\ra ~=~-{1\o 2}\ln |w-z|~~,}
so that in Minkowski space $\varphi=\Phi/\sqrt{\pi}$ is
described by the usual lagrangian ${1\o 2}\partial_\mu \varphi \partial^\mu
\varphi$.
The correlation functions of the chiral fields $\phi(z)$ and $\phib(\zb)$
in the decomposition  $\Phi(z,\zb)={1\o 2}(\phi(z) +\phib(\zb))$ ~are~
$\la \phi(w)\phi(z)\ra=-\ln(w-z)$ ~and~
$\la\phib(\wb)\phib(\zb)\ra=-\ln(\wb-\zb)$.

The chiral fields have
the mode expansion
\eqn\phimodes{\phi(z)= q-i\alp_0 \ln z +i \sum_{n\neq 0} {1\o n}\alp_n z^{-n}~,
  ~~ \phib(\zb) = \bar{q}-i \bar{\alp}_0 \ln\zb +
    i \sum_{n\neq 0} {1\o n}\bar{\alp}_n \zb^{-n}~,}
where the nonvanishing commutators of the operators $q, ~\alp_n$ and
$\bar{q}, ~\bar{\alp}_n$ are
\eqn\CRs{ [q,\alp_0] ~=~ [\bar{q},\bar{\alp}_0] ~=~ i ~,~~~~
[\alp_n,\alp_m] ~=~ [\bar{\alp}_n,\bar{\alp}_m] ~=~ n \del_{n+m,0} ~.}
Since all left- and right-moving fields commute, the normal ordering of the
$V_{m,n}$ in~\Vmn\ follows from that of a chiral ``half''
\eqn\NO{ :e^{\phi(z)}:
  ~\equiv~ e^{\phi_-(z)}~e^q z^{-i \alp_0}~e^{\phi_+(z)}~, }
where $\phi_\pm(z)$ is the part of $\phi(z)$ that contains only modes with
$n>0$, respectively, $n<0$.

\appendix{B}{The Statistical Mechanics Connection}

Taking the scaling limit of certain types of one-dimensional
quantum spin chains
(or various two-dimensional lattice models)
is often regarded as {\it the} way to rigorously define
and study (1+1)-dimensional relativistic QFTs.
In particular, the spin ${1\o 2}$ XYZ Heisenberg chain is
often
 mentioned
as an appropriate regularized system
whose scaling limit defines
the SGM.
We will here examine this issue slightly more
carefully,
though still at a rather heuristic level.

Consider the hamiltonian
\eqn\chain{ H(\gam,\veps,\vec{h})~=~
  \sum_{n=1}^N [    (1+\veps)
  \sig_n^x \sig_{n+1}^x + (1    -    \veps)\sig_n^y \sig_{n+1}^y +
   \cos\gam ~\sig_n^z \sig_{n+1}^z + \vec{h}\cdot\vec{\sig}_n ]~~,}
where the $\vec{\sig}_n$ are Pauli matrices at each site of the chain.
Throughout the discussion below $N$ is taken to be even, $0\leq \gam<\pi$, and
periodic boundary conditions $\vec{\sig}_{\sN+1}=\vec{\sig}_1$ are imposed.
$H(\gam,0,0)$ is then the
(antiferromagnetic) XXZ chain, whose scaling limit is known~\rXXZi\ to
be described by the {\it bosonic} gaussian CFT at
$r=r(\gam)=[2(1-\gam/\pi)]^{-1/2}$
(or its dual).
The electric charge $m$ of states in the CFT
is just the conserved total spin $S^z={1\o 2}\sum_n \sig_n^z$
in the $z$-direction
(which takes integer values when $N$ is even).
The additional $U(1)$ symmetry of the CFT, corresponding to the magnetic
charge $n$, is not manifest in the XXZ hamiltonian.

Now, based on the known~\rXXZii\
large-distance behaviour of
certain correlation functions in the XXZ
antiferromagnet one identifies the CFT operators
$V_{\pm 1,0}$ as the continuum versions of
$\sig_n^\pm \equiv \sig_n^x \pm i\sig_n^y$, respectively (up to
symmetry
transformations and possibly subleading irrelevant operators). We are therefore
led to identify the scaling limit of the XXZ chain in a
transverse field,
$H(\gam,0,h\hat{n}_\perp)$ with $h\to 0$
(where $\hat{n}_\perp \cdot \hat{n}_z=0$), as being described by
$A_{\rb}(r(\gam),V_{1,0}^{(+)})=$~SG$(\beta \! =\! \sqrt{2(\pi -\gam)})$.

This
claim may sound surprising at first,
since it is
believed
  (see \eg~the first reference in~\rXXZi)
that the XXZ chain in a transverse field is not integrable
(contrary to the longitudinal case).
However, such a situation would not be new:
The Ising model in a magnetic field
is apparently non-integrable on the lattice,      but
its scaling limit (at $T=T_c$) is conjectured to be
an integrable QFT. Evidence supporting the claim in the Ising case
were provided~\rIsnum\ by numerical simulations of the corresponding chain
and lattice models, and we think it is worthwhile to perform similar checks
of our prediction in the case of the XXZ chain
in a transverse field.

Next, consider the scaling limit of the XYZ chain in the absence of
a magnetic field, namely $H(\gam,\veps,0)$
with $\veps \to 0$, which differs from the XXZ
hamiltonian by a $\sum_n (\sig_n^+ \sig_{n+1}^+ + \sig_n^- \sig_{n+1}^-)$
``perturbation''. It is now
natural to conjecture    that it is
described by the perturbed CFT $A_{\rb}(r(\gam),V_{2,0}^{(+)})$,
when $0<\gam <\pi$
(cf.~\rHenSal).
This theory is identified (see~\rourviii\ and sect.~3.1) as
SG($\beta=\sqrt{8(\pi-\gam)}$,2), namely a ``massive orbifold'' of
SG($\beta=\sqrt{8(\pi-\gam)}$) in which the vacuum is doubly-degenerate.
This latter fact is consistent with the asymptotic double degeneracy
of the XYZ chain ground state energy   in the thermodynamic limit~\rBaxt.

Finally, let us comment about
the relation between the XYZ hamiltonian and the lattice MTM.
The MTM lattice hamiltonian can be obtained~\rLuth\
from that of the XYZ chain by a Jordan-Wigner transformation. However, the
boundary conditions on the resulting fermion operators have
 to be handled
with care. Exactly like in the case of the Ising model versus the lattice
free Majorana fermion, in order
to preserve the original boundary conditions that were
imposed on the bosonic variables of the XYZ chain
one has to consider ``simultaneously''
different boundary conditions on the fermions~\rLuMTM. This is precisely the
lattice analog of the GSO projection, or ``twist'' with respect to the
fermion number (see sect.~2), that goes back to~\rKauf\ in the Ising case.
The continuum MTM is obtained by taking the scaling limit of the Jordan-Wigner
transformed XYZ chain
with {\it one} choice of boundary conditions
on the fermions (imposing anti-periodic
boundary conditions one probes
the full mutually-local operator algebra of the theory).
In the perturbed CFT language, this procedure of ``undoing'' the GSO
projection takes us from $A_{\rb}(r(\gam),V_{2,0}^{(+)})$
to $A_{\rf}(r(\gam),V_{2,0}^{(+)})$; the latter being the MTM is
in agreement with the discussion of sect.~3.

\vfill\eject
\listrefs

\bye\end